\documentclass[USenglish,a4paper,11pt]{article}
\usepackage{babel}
\usepackage{microtype}
\usepackage{amsmath,amsthm,amsfonts,amssymb}
\usepackage{mathtools}
\usepackage{thmtools,thm-restate}
\usepackage{nicefrac}
\usepackage{graphicx}
\usepackage{booktabs}
\usepackage{multirow}
\usepackage{titling}
\usepackage{authblk}
\usepackage{xcolor}
\usepackage[colorlinks=true,linkcolor=blue,citecolor=teal]{hyperref}
\usepackage{fullpage}
\usepackage[sort&compress,numbers]{natbib}
\usepackage[nameinlink]{cleveref}
\usepackage{float}
\usepackage{placeins}
\usepackage{appendix}
\usepackage{tocloft}
\usepackage{tikz}
\usetikzlibrary{arrows.meta, positioning, calc}
\usetikzlibrary{patterns}
\usetikzlibrary{shapes.geometric}
\newcommand{\bigO}{\mathcal{O}}
\usepackage[linesnumbered,ruled,vlined]{algorithm2e}
\usepackage{xurl}

\newtheorem{theorem}{Theorem}
\newtheorem{lemma}[theorem]{Lemma}

\newtheorem{definition}[theorem]{Definition}

\usepackage[utf8]{inputenc}
\usepackage[T1]{fontenc}
\usepackage{caption}
\usepackage{subcaption}
\usepackage[scr=boondox]{mathalfa}
\usepackage{enumitem}

\usepackage{csquotes}

\DeclareMathOperator*{\argmax}{arg\,max}

\usepackage{complexity}

\graphicspath{{./figures/}}

\makeatletter
\AddToHook{cmd/appendix/before}{\def\cref@section@alias{appendix}}
\AddToHook{cmd/appendix/before}{\def\cref@subsection@alias{appendix}}
\makeatother

\title{What to Predict for Efficient Scheduling on Parallel Machines}

\newcommand{\email}[1]{\texttt{#1}}

\author[1]{Evripidis Bampis}
\author[1]{Bruno Escoffier}

\affil[1]{Sorbonne Universit\'e, CNRS, LIP6, F-75005 Paris, France
\authorcr
\email{\string{evripidis.bampis, bruno.escoffier, georgios.mitropoulos\string}@lip6.fr}}

\author[2,3]{Dimitris Fotakis}
\author[1]{Giorgos Mitropoulos}

\affil[2]{National Technical University of Athens, Greece}
\affil[3]{Archimedes, Athena Research Center, Greece
\authorcr
\email{fotakis@cs.ntua.gr}}

\author[4]{Michalis Xefteris}

\affil[4]{University of Bremen, Germany
\authorcr
\email{xefteris@uni-bremen.de}}

\begin{document}
\maketitle
\begin{abstract}
We study learning-augmented algorithms for scheduling on multiple parallel machines under the objectives of minimizing the makespan or the total weighted completion time. We assume that the algorithm has access to noisy information (the prediction) about an optimal solution, expressed through a solution encoding that implicitly quantifies both the amount of information provided and the nature of noise (errors in the prediction). Our central question is to determine the maximum level of noise that still allows for efficiently recovering this optimal solution.
We observe that information guiding greedy scheduling to optimality or revealing the optimal assignment of tasks to machines is very sensitive to noise, and prove that recovering an optimal solution from solution encodings capturing either is computationally hard, even for a small (non-constant) number of errors. On the positive side, we show that an order-based encoding, which leads to errors that are dispersed locally, is robust to noise. We present a dynamic programming framework that utilizes this encoding and solves optimally several classical scheduling problems, including makespan and total weighted completion time minimization on identical or unrelated machines.
\end{abstract}

\section{Introduction}

Modern breakthroughs in machine learning have motivated researchers to incorporate \emph{predictions} of unknown quality into algorithm design for optimization problems, giving rise to the rapidly growing field of \emph{algorithms with predictions}, also known as \emph{learning-augmented algorithms}. 
The goal is to use such predictions to improve algorithmic 
performance when they are accurate, while still retaining rigorous guarantees when they are not. 
This line of research originated in online optimization, starting with the work of Mahdian et al.~\cite{mahdian} on online advertising allocation, and was later formalized by Lykouris and Vassilvitskii~\cite{lykouris}.
Since then, this framework has been applied to~various problems, including online algorithms, data structures, streaming, and combinatorial optimization.
An extensive list of works in this area is maintained in the algorithms-with-predictions 
repository~\cite{alps}.

More recently, predictions have also been used to overcome the computational challenges of \NP-hard problems, leading to algorithms with improved approximation guarantees for several classical optimization problems.
One of the main directions in this line of work is clustering. 
For $k$-means,~several learning-augmented algorithms use predicted cluster labels and achieve a $(1+\bigO(\alpha))$-approximation when the per-cluster label error is at most $\alpha$~\cite{clustering1, clustering2, clustering3}. 
Hierarchical clustering has also been studied in this framework~\cite{hierarchical}. 
Beyond clustering, Antoniadis et al.~\cite{antoniadis} considered a broad class of \NP-hard selection problems and showed how predictions can lead to improved approximation guarantees when they are sufficiently accurate. 
Another active direction concerns Max-CSP problems, and in particular Max-Cut. 
Bampis et al.~\cite{maxcut1} showed that predictions can be used to speed up polynomial-time approximation schemes for dense Max-CSPs. 
For Max-Cut, Cohen-Addad et al.~\cite{maxcut2} and Ghoshal et al.~\cite{maxcut3} independently studied a model in which each prediction is correct with probability slightly larger than $1/2$, obtaining improved approximation ratios. 
The same type of prediction has also been applied to other \NP-hard problems~\cite{independentset, graphapprox}, giving guarantees beyond those known without predictions.

Learning-augmented algorithms for scheduling have been studied mostly in online settings~\cite{online1, online2}. 
The offline setting is less explored. 
Moseley et al.~\cite{stochscheduling} studied stochastic single-machine scheduling, where the scheduler is given predicted job-size distributions that may only approximate the true distributions. 
Closer to our work, Bampis et al.~\cite{icml_paper} built on the noisy-sorting framework of Braverman and Mossel~\cite{noisy_sorting} to study permutation optimization problems with predictions. 
In their setting, the prediction is a noisy version of an optimal ordering, and the goal is to recover an exact optimal solution. 
They showed that, for a broad class of permutation problems satisfying suitable structural properties, an optimal solution can be recovered with high probability in polynomial time. 
A notable application is $1\mid \mathrm{prec}\mid \sum C_j$, the single-machine problem of minimizing the sum of completion times subject to precedence constraints.\footnote{We use the standard three-field notation $\alpha\mid\beta\mid\gamma$ for scheduling problems~\cite{3-field}, where $\alpha$ denotes the machine environment, $\beta$ the job constraints, and $\gamma$ the objective. We recall this notation in Section~\ref{sec:preliminaries}.} 
To our knowledge, this was the first use of predictions to obtain in polynomial time an exact solution for an \NP-hard scheduling problem, and more generally for an \NP-hard problem, in the offline setting.

Motivated by the results of Bampis et al.~\cite{icml_paper}, we aim to understand the following question: suppose that a prediction oracle provides a noisy encoding of information about an optimal solution for some \NP-hard problem. What kind of information about the optimal solution, what kind of noise and how much of it, can be tolerated so that an optimal solution can still be recovered  
in polynomial time? Our prediction model is conceptually related to advice complexity for online algorithms~\cite{boyar_advice_survey} and untrusted advice~\cite{angelopoulos_untrusted_advice}. In the online setting, advice helps overcome uncertainty about future inputs. Here, the full input is known, and predictions help overcome computational hardness.

Previous work on learning-augmented algorithms for \NP-hard problems typically fixes a particular form of prediction and does not investigate how different kinds of noisy information about an optimal solution affect its usefulness towards recovering the initial solution.
In this work, we initiate a systematic study of this question for scheduling problems on parallel machines. 
We focus on makespan minimization and total weighted completion time involving $n$ jobs and multiple machines, where the number of machines may be fixed or part of the input. We study several natural ways of encoding a predicted schedule and ask whether the corresponding prediction contains enough information to recover an optimal solution when it is noisy. 
Our results show that most of these encodings are highly sensitive to prediction errors and do not allow for efficient exact recovery when the number of errors is non-constant, while only one encoding contains enough information to recover optimal schedules in several scheduling settings.

\smallskip\noindent\textbf{Our Contribution.}
First, we consider a setting where the predictor provides a job ordering such that assigning jobs sequentially to the currently least-loaded machine yields an optimal makespan. 
We call this a \emph{greedy-based encoding} of a solution. 
For this encoding, we prove that even if each job in the predicted ordering is displaced by at most one position relative to an optimal greedy ordering, $P2\mid\mid C_{\max}$ remains \NP-complete. 
This intractability suggests that greedy-based encodings are not robust enough to serve as a prediction model for exact recovery in practice.

Second, we assume that we are given a predicted assignment of jobs to machines, which we call a \emph{partition-based encoding} of a schedule. 
For this encoding, we prove that if at most $k$ jobs are assigned to the wrong machine, then recovering the optimal solution for $P2\mid\mid C_{\max}$ cannot be achieved in time $n^{o(k)}$ under the Exponential Time Hypothesis (ETH). 
Thus, this prediction model can lead to exact polynomial-time algorithms only when the number of incorrectly assigned jobs is constant.

Next, we investigate whether explicitly identifying the prediction errors can help overcome these limitations. We equip the partition-based encoding with error-awareness, resulting in an \emph{error-aware partition-based encoding}, where the given assignment of jobs to machines comes with the set of jobs that may be misplaced. For a fixed number of machines $m$, this information allows one to solve $Pm\mid\mid C_{\max}$ in polynomial time when at most $\bigO(\log n)$ jobs are flagged on each machine, by brute-forcing the possible destinations of all flagged jobs. 
We show, however, that when $m$ is part of the input, this encoding still fails to provide sufficient information for exact recovery: $P\mid\mid C_{\max}$ remains strongly $\NP$-complete even if at most three jobs are flagged on each machine.

The main limitation of the error-aware partition-based encoding is that it does not hint to the machine where a misplaced job needs to be reassigned. To address this, we introduce a locality condition on prediction errors. 
Given $m$ indexed machines, a schedule can be viewed as an $m$-tuple of sequences, where each sequence contains the jobs scheduled on one machine in their processing order. Our prediction is simply this $m$-tuple presented as a one-dimensional sequence, by listing the sequence of jobs on machine~$1$, followed by those on machine~$2$, and so on.
We call this an \emph{order-based encoding} of a solution. 
This encoding allows both errors in the order of jobs on a machine and local errors in the machine assignment. In the single-machine setting, this encoding coincides with the predicted ordering considered by Bampis et al.~\cite{icml_paper}.
Although this prediction may seem strong, it can be obtained from independent noisy pairwise comparisons between jobs. 
Indeed, by~\cite{noisy_sorting}, if each pairwise comparison is correct with probability at least $1/2+\varepsilon$, for some constant $\varepsilon>0$, with respect to the true lexicographic order of the jobs, then with high probability one can recover in polynomial time an order-based encoding in which each job is displaced from its optimal position by at most $k=\bigO(\log n)$. 
Here, the lexicographic order first sorts jobs by the index of the machine on which they are scheduled and then by their position on that machine.

We show that the order-based encoding contains enough information to efficiently recover optimal solutions for several scheduling problems whose objective value is determined by the partition of jobs among the machines. For objectives of the form $\max_{i\in[m]} f_i(S_i)$, where $S_i$ denotes the set of jobs assigned to machine $i$, we give a dynamic programming (DP) algorithm that is fixed-parameter tractable in $k$ (and polynomial for $k=\bigO(\log n)$) whenever the functions $f_i$ can be evaluated in polynomial time. The idea is that a prefix of the hidden optimal order corresponds to the set of jobs assigned to the first machines, and the predicted order determines this set up to a small number of locally ambiguous jobs. The dynamic program enumerates these possibilities and recovers the optimal partition. 
This framework captures, in particular, $P\mid\mid C_{\max}$, as well as makespan minimization on unrelated machines. 
It also extends to weighted sum of completion times, and more generally all objectives of the form $\sum_{i\in[m]} f_i(S_i)$, whenever the corresponding single-machine value $f_i(S_i)$ can be computed efficiently after the set $S_i$ is fixed.
We also study scheduling with release times. 
Makespan minimization with release times fits directly into the same framework, since the optimal single-machine makespan can be computed in polynomial time once the assigned set of jobs is fixed. 
For the more challenging objective $P\mid r_j\mid \sum w_j C_j$, the single-machine problem induced by a fixed partition is already \NP-hard~\cite{lenstra}. We show that the order-based prediction allows us to solve this single-machine subproblem in time fixed-parameter tractable in $k$, and we combine this subroutine with the previous dynamic program to obtain an algorithm for the parallel-machine problem.

Finally, we consider precedence constraints. 
For a constant number of machines, we obtain a pseudopolynomial-time algorithm when $k=\bigO(\log n)$ for $Pm\mid \mathrm{prec}\mid C_{\max}$. 
The algorithm first enumerates the possible assignments of jobs to machines, consistently with the predicted order, and then uses a dynamic programming table to sequentially determine the position of each job on its assigned machine while respecting the precedence constraints. Again, this extends to different objective functions, such as sum of weighted completion times.

We also address the limitations of such an order-based prediction and show that, although it is powerful, it is not a silver bullet for all scheduling problems.
Specifically, we prove that it cannot lead to a polynomial-time algorithm for $P\mid \mathrm{prec}\mid C_{\max}$ when the number of machines is part of the input, unless $\P=\NP$. 
This hardness result is obtained via a reduction from $3$-SAT and holds even when every job in the predicted ordering is at most two positions away from its position in an optimal ordering.
Table~\ref{tab:main_results} summarizes our main results for the makespan objective.

\begin{table}[h!]
\centering
\small
\setlength{\tabcolsep}{3pt}
\renewcommand{\arraystretch}{1.2}
\begin{tabular}{c||c|c|c}
Encoding model and error 
& $P\mid\mid C_{\max}$ 
& $Pm\mid \mathrm{prec}\mid C_{\max}$ 
& $P\mid \mathrm{prec}\mid C_{\max}$ \\
\hline
\hline
Greedy-based encoding
& \multicolumn{3}{c}{\NP-hard for $k=1$, Thm.~\ref{thrm:greedy_hard}} \\
$k=$ max displacement
& \multicolumn{3}{c}{(even for $P2\mid\mid C_{\max}$)} \\
\hline
Partition-based encoding
& \multicolumn{3}{c}{No $n^{o(k)}$-time algo under ETH, Thm.~\ref{thrm:partition_hard}} \\
$k=$ \# of misplaced jobs
& \multicolumn{3}{c}{(even for $P2\mid\mid C_{\max}$)} \\
\hline
Order-based encoding
& Polynomial if
& Pseudopolynomial if
& \NP-hard for $k=2$ \\
$k=$ max displacement
& $k=\bigO(\log n)$, Thm.~\ref{thrm:outer_dp}
& $k=\bigO(\log n)$, Thm.~\ref{thrm:order_prec_dp}
& Thm.~\ref{thrm:hard_order-based} \\
\hline
\end{tabular}
\vspace{.5em}
\caption{Summary of results for $C_{\max}$. Tractability results also apply to other scheduling problems.}
\label{tab:main_results}
\end{table}

The aforementioned learning-augmented algorithms can also be made robust to inaccurate predictions. If a polynomial-time $\rho$-approximation algorithm is available for the same problem without predictions, we can run it in parallel with our DP using a chosen parameter $K=\bigO(\log n)$. After both algorithms terminate, we return the best feasible schedule found. This gives a $\rho$-approximation for arbitrary predictions (robustness) and an optimal schedule whenever the prediction is a $K$-perturbed order-based encoding of an optimal solution.

The paper is organized as follows: Section~\ref{sec:preliminaries} introduces notation and encodings. Section~\ref{sec:hardness} presents hardness results for various prediction types, while Section~\ref{sec:outer_DP} develops a DP framework for order-based predictions and release times. We address precedence constraints in Section~\ref{sec:order_prec}, explore limitations of order-based predictions in Section~\ref{sec:limitations_order}, and conclude in Section~\ref{sec:conclusion}. 
Due to space constraints, some proofs are deferred to the appendix.

\section{Preliminaries}
\label{sec:preliminaries}

Throughout this work, we use the standard three-field notation $\alpha\mid\beta\mid\gamma$ for scheduling problems~\cite{3-field}. 
Here, $1$ denotes a single machine, $P$ denotes identical parallel machines when the number of machines is part of the input, and $Pm$ denotes identical parallel machines when the number $m$ of machines is fixed. 
An empty second field means that there are no additional job constraints, while $r_j$ denotes release times, meaning that job $j$ cannot start before time $r_j$, and $\mathrm{prec}$ denotes precedence constraints. 
In the third field, $C_{\max}$ denotes the makespan, while $\sum C_j$ and $\sum w_j C_j$ denote the total and weighted total completion time, respectively.

\begin{definition}[$P \mid \mathrm{prec} \mid C_{\max}$]
    Let $J=\{j_1,\ldots,j_n\}$ be a set of jobs with respective processing times $p_1,\ldots,p_n$, and let $G_{p}=(J,E_p)$ be a directed acyclic graph (DAG) of precedence constraints among the jobs. Given $m$ identical parallel machines, find a feasible schedule $S$ that assigns jobs to machines and schedules them on each machine, respects the precedence constraints, and minimizes the makespan $C_{\max}$. If there are no precedence constraints, the problem is denoted by $P\mid\mid C_{\max}$.
\end{definition}

In this paper, we often encode solutions as orderings of jobs $j_1, \dots, j_n$. We denote by $\sigma$ the job ordering $(j_{\sigma(1)},\dots, j_{\sigma(n)})$. For such an ordering $\sigma$, we write $\sigma^{-1}(a)$ for the position of job $j_a$ in $\sigma$. We also use the notation $[n] = \{1,\dots,n\}$ and $[i,j] = \{i,\dots,j\}$. Throughout this paper, we often refer to the time window $(i-1, i]$, for $i \in  \mathbb{Z}^+$, simply as \emph{time slot}~$i$. Thus, a single job with processing time greater than $1$ may occupy more than one time slot. In the case of multiple parallel machines, there exist as many time slots~$i$ as the number of machines.

\begin{definition}[Solution encodings]
    Given an instance of a scheduling problem, let $S$ be a (potentially infeasible) schedule. The schedule $S$ can be encoded in the following natural ways:
    \begin{enumerate}

        \item 
        \emph{Greedy-based encoding:} A solution is represented by an ordering of jobs. The schedule follows by sequentially assigning each job to the machine that first becomes available.

        \item 
        \emph{Partition-based encoding:} A solution is represented by $m$ subsets $S_1,\dots,S_m$ that form a partition of jobs to machines. When the objective function of the scheduling problem at hand depends on the order of jobs within each machine, an ordering of jobs in each set is also provided.

        \item 
        \emph{Order-based encoding:} A solution is represented by a one-dimensional array obtained by concatenating the ordering of jobs within each machine in sequential order (i.e. the jobs of machine~$1$, followed by those of machine~$2$, etc.). 
        \end{enumerate}
\end{definition}

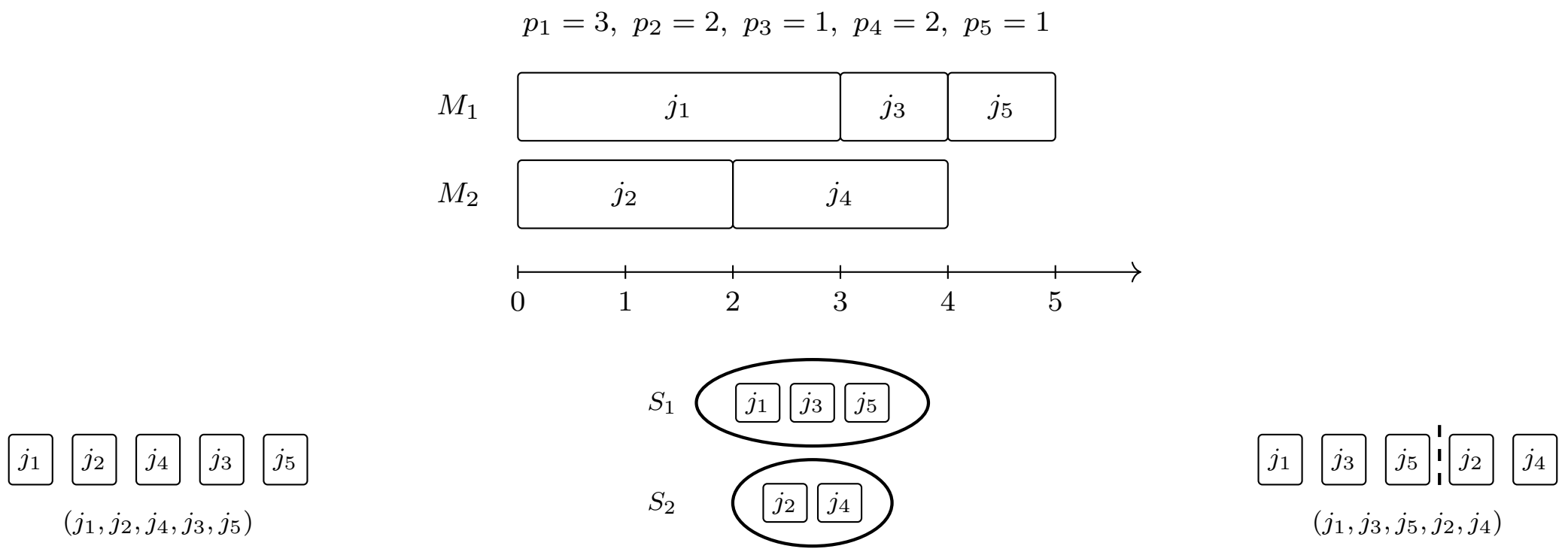
\begin{figure}[H]
    \centering
    \begin{tikzpicture}[
        scale=0.9,
        transform shape,
        x=1.05cm,
        y=0.95cm,
        job/.style={draw, rounded corners=1pt, minimum height=0.55cm, inner sep=1pt, font=\scriptsize, align=center},
        mach/.style={font=\scriptsize\bfseries},
        tick/.style={font=\scriptsize}
    ]
        \node[mach] at (-0.55,0.35) {$M_1$};
        \node[mach] at (-0.55,-0.55) {$M_2$};

        \draw[job] (0,0) rectangle (3,0.7) node[midway] {$j_1$};
        \draw[job] (3,0) rectangle (4,0.7) node[midway] {$j_3$};
        \draw[job] (4,0) rectangle (5,0.7) node[midway] {$j_5$};

        \draw[job] (0,-0.9) rectangle (2,-0.2) node[midway] {$j_2$};
        \draw[job] (2,-0.9) rectangle (4,-0.2) node[midway] {$j_4$};

        \draw[->] (0,-1.35) -- (5.8,-1.35);
        \foreach \x in {0,1,2,3,4,5} {
            \draw (\x,-1.28) -- (\x,-1.42);
            \node[tick] at (\x,-1.65) {$\x$};
        }

        \node[font=\scriptsize] at (2.5,1.2)
        {$p_1=3,\ p_2=2,\ p_3=1,\ p_4=2,\ p_5=1$};
    \end{tikzpicture}

    \vspace{0.7em}

\begin{subfigure}{0.31\textwidth}
    \centering
    \begin{tikzpicture}[
        scale=0.8,
        transform shape,
        job/.style={draw, rounded corners=1pt, minimum height=0.55cm, minimum width=0.48cm, inner sep=1pt, font=\scriptsize}
    ]
        \def\jobsep{0.70}

        \node[job] at (0,0) {$j_1$};
        \node[job] at ({\jobsep},0) {$j_2$};
        \node[job] at ({2*\jobsep},0) {$j_4$};
        \node[job] at ({3*\jobsep},0) {$j_3$};
        \node[job] at ({4*\jobsep},0) {$j_5$};

        \node[font=\scriptsize] at ({2*\jobsep},-0.7)
        {$(j_1,j_2,j_4,j_3,j_5)$};
    \end{tikzpicture}
    \caption{Greedy-based encoding}
\end{subfigure}
\hfill
\begin{subfigure}{0.31\textwidth}
    \centering
    \begin{tikzpicture}[
        scale=0.8,
        transform shape,
        job/.style={draw, rounded corners=1pt, minimum height=0.42cm, minimum width=0.48cm, inner sep=1pt, font=\scriptsize},
        setone/.style={draw, ellipse, minimum width=2.55cm, minimum height=0.95cm, thick},
        settwo/.style={draw, ellipse, minimum width=1.75cm, minimum height=0.95cm, thick},
        label/.style={font=\scriptsize\bfseries}
    ]
        \def\jobsep{0.70}

        \node[setone] at (0,0.45) {};
        \node[label] at (-1.65,0.45) {$S_1$};
        \node[job] at ({-0.6},0.45) {$j_1$};
        \node[job] at (0,0.45) {$j_3$};
        \node[job] at ({0.6},0.45) {$j_5$};

        \node[settwo] at (0,-0.65) {};
        \node[label] at (-1.65,-0.65) {$S_2$};
        \node[job] at ({-0.6/2},-0.65) {$j_2$};
        \node[job] at ({0.6/2},-0.65) {$j_4$};
    \end{tikzpicture}
    \caption{Partition-based encoding}
\end{subfigure}
\hfill
\begin{subfigure}{0.31\textwidth}
    \centering
    \begin{tikzpicture}[
        scale=0.8,
        transform shape,
        job/.style={draw, rounded corners=1pt, minimum height=0.55cm, minimum width=0.48cm, inner sep=1pt, font=\scriptsize},
        cut/.style={dashed, thick}
    ]
        \def\jobsep{0.70}

        \node[job] at (0,0) {$j_1$};
        \node[job] at ({\jobsep},0) {$j_3$};
        \node[job] at ({2*\jobsep},0) {$j_5$};

        \draw[cut] ({2.5*\jobsep},0.38) -- ({2.5*\jobsep},-0.38);

        \node[job] at ({3*\jobsep},0) {$j_2$};
        \node[job] at ({4*\jobsep},0) {$j_4$};

        \node[font=\scriptsize] at ({2*\jobsep},-0.7)
        {$(j_1,j_3,j_5,j_2,j_4)$};
    \end{tikzpicture}
    \caption{Order-based encoding}
\end{subfigure}

    \caption{A schedule on two machines and five jobs, with the three encodings of the same solution.}
    \label{fig:solution_encodings}
\end{figure}

Figure~\ref{fig:solution_encodings} illustrates these three encodings on the same schedule with two machines and five jobs.

\section{Hardness results for different prediction models}
\label{sec:hardness}
In this section, we show hardness results when using predictions based on different solution encodings.
In particular, we first prove the greedy order is very noise sensitive, to the point where recovering the optimal makespan from an almost optimal prediction remains \NP-complete. We then move to partition-based predictions, where we show that simply knowing an assignment of jobs to machines does not suffice to efficiently recover a makespan minimizing schedule, even when we know exactly which jobs are misplaced. Missing proofs can be found in Appendix~\ref{app:hardness}.

\subsection{Predictions on the greedy order for makespan minimization}

Let $\sigma^*$ be an ordering of jobs such that a greedy criterion\footnote{In makespan minimization, the standard greedy algorithm assigns the next job in the order to the least loaded machine.} is guaranteed to produce an optimal assignment. We call this an \emph{optimal greedy ordering}. We now assume that we have access to a predicted greedy ordering $\sigma$ with respect to $\sigma^*$.

\begin{definition}[$Pm \mid \mid C_{\max}$ with $k$-perturbed greedy-based encoding]
    Let $I$ be an instance of $Pm \mid \mid C_{\max}$ with a perturbed greedy ordering $\sigma$ of an optimal greedy ordering $\sigma^*$ that minimizes makespan, and $t$ be the target makespan. If $|\sigma^{-1}(i)-(\sigma^*)^{-1}(i)|\leq k$, for all $i \in [n]$, is there an $m$-way partition of jobs to machines with makespan at most $t$?
\end{definition}

We prove that even if each job in $\sigma$ is at most {\it one} position away from its position in $\sigma^*$, $Pm \mid \mid C_{\max}$ remains \NP-complete. This is shown via a reduction from $Pm \mid \mid C_{\max}$ without any information.

\begin{restatable}{theorem}{thmgreedyhard}
\label{thrm:greedy_hard}
    $Pm \mid \mid C_{\max}$ with a $1$-perturbed greedy order is weakly $\NP$-complete, even when $m=2$.
\end{restatable}

\subsection{Partition-based predictions for makespan minimization}

In this subsection, we assume that we have access to a predicted assignment of all jobs to machines. We prove that if this predicted partition differs from some optimal one in at most $k$ jobs, then $Pm \mid \mid C_{\max}$ cannot be solved in $n^{o(k)}$ time, under ETH.

\begin{definition}[$Pm \mid \mid C_{\max}$ with $k$-perturbed partition-based encoding]
    Let~$I$ be an instance of $Pm \mid \mid C_{\max}$ and $t$ be a target makespan. Let $S_1,...,S_m$ be a perturbed partition such that there exists a makespan-minimizing partition $S_1^*,...,S_m^*$ with $|S_1 \setminus S_1^*|+...+|S_m \setminus S_m^*| \leq k$. Is there an $m$-way partition of jobs to machines with makespan at most $t$?
\end{definition}

We get this hardness result via a reduction from $d$-SUM. Intuitively, the difficulty of recovering the optimal solution stems from the fact that knowing only the number of erroneous assignments provides no information about which specific jobs are assigned to the wrong machine, thus creating a computational challenge that is hard to overcome.

\begin{definition}[$d$-SUM]
    Given a (multi)set $A$ of $n$ integers, is there a subset of $A$ with cardinality~$d$ whose elements sum to zero?
\end{definition}

\begin{theorem}[\cite{kSUM}] 
    Under ETH, for every $d < n^{0.99}$, $d$-SUM over $n$~numbers of $\bigO(d \log n)$ bits cannot be solved in $n^{o(d)}$ time.
\end{theorem}

\begin{theorem} \label{thrm:partition_hard}
    Under ETH, $Pm \mid \mid C_{\max}$ with $k$-perturbed partition-based encoding is not solvable in $n^{o(k)}+\Theta(n)$ time, uniformly for any $k<n^{0.99}$.
\end{theorem}

\begin{proof}
    Consider an instance $A=\{a_1,...,a_n\}$ of $k$-SUM. Let $B = 1+2\sum_{i=1}^n |a_i|$ and $A' = \{a + B \mid a \in A\}$. It follows directly that $A$ is a yes-instance for $k$-SUM if and only if there exists a subset of $A'$ with cardinality $k$ that sums up to $kB$.

    We transform $A'$ into an instance $I'$ of $Pm \mid \mid C_{\max}$ with $k$-perturbed partition-based encoding. The instance consists of a set of $n+1$ jobs: jobs $\{j_1,...,j_n\}$ correspond to the elements of $A'$, with each job $j_i$, $i \in [n]$, having processing time equal to $a_i+B$, and job $j_{n+1}$ has processing time $\sum_{i=1}^n (a_i + B) - 2kB$ (assuming $k \leq n/2 - 1$). It follows directly that the minimum makespan for this instance should be at least $\sum_{i=1}^n (a_i + B) - kB$; we set this as our target $t$. The predicted partition consists of the subsets $S_1 = \{j_1,...,j_n\}$ and $S_2=\{j_{n+1}\}$. 

    We now show that a makespan of $\sum_{i=1}^n (a_i + B) - kB$ can be achieved on instance $I'$ if and only if a witness can be found by moving exactly $k$ jobs from $S_1$ to $S_2$. This allows us to establish two crucial facts: first, moving $p>k$ items to the set that contains $j_{n+1}$ leads to a strictly greater makespan, meaning that there must exist a makespan-minimizing partition $S_1^*,\ S_2^*$ such that $|S_1\setminus S_1^*| + |S_2 \setminus S_2^*| \leq k$ holds (so we have a valid input for our target problem). Second, a schedule showing that we have a yes-instance of $P2 \mid \mid C_{\max}$ needs to move exactly $k$ jobs with $j_{n+1}$.
    
    Scheduling any subset $U$ of $k$ jobs in the same machine as job $j_{n+1}$ would lead to a makespan of
    \begin{align}
    \begin{split}
        &\max \left\{\sum_{i=1}^n (a_i + B) - \sum_{i \in U} (a_i+B),\  \sum_{i=1}^n (a_i + B) - 2kB + \sum_{i \in U} (a_i+B) \right\} \\
        &=\max \left\{\sum_{i=1}^n (a_i + B) - kB-\sum_{i \in U} a_i,\  \sum_{i=1}^n (a_i + B) - kB + \sum_{i \in U} a_i \right\} \\
        &= \sum_{i=1}^n (a_i + B) -kB +\Big|\sum_{i\in U} a_i\Big| \leq \sum_{i=1}^n (a_i + B) -kB +\sum_{i=1}^n |a_i|.  
    \end{split}\label{eq:partition_makespan_ub}
    \end{align}

    If we schedule a set $U'$ of $p>k$ jobs together with $j_{n+1}$, then the machine with $j_{n+1}$ would have makespan at least $\sum_{i=1}^n (a_i+B)-2kB+pB+\sum_{i\in U'}a_i\geq \sum_{i=1}^n (a_i+B)-kB+B-\sum_{i=1}^n |a_i|$. This is strictly greater than the upper bound obtained when we move $k$ jobs (as $B>2\sum_{i=1}^n |a_i|$). Similarly, if we schedule a set $U'$ of $p<k$ jobs together with $j_{n+1}$, then the machine without $j_{n+1}$ would have makespan at least $\sum_{i=1}^n (a_i+B) - \sum_{i\in U'}(a_i+B)\geq \sum_{i=1}^n (a_i+B) - kB + B -\sum_{i=1}^n |a_i|$, again strictly greater than the upper bound obtained when we move $k$ jobs. 
    
    Therefore, this is a yes-instance of $P2 \mid \mid C_{\max}$ with $k$-perturbed partition-based encoding with target $\sum_{i=1}^n (a_i + B) - kB$ if and only if there exists a subset $U$ of $k$ jobs with total processing time equal to~$kB$, meaning $\sum_{i\in U} a_i=0$, i.e., if and only if $A$ is a yes-instance for $k$-SUM.
    
    Suppose that we are able to answer this problem in $n^{o(k)}+\Theta(n)$ time.
    Both the construction of $A'$ and the transformation of $A'$ into the instance $I'$ require $\bigO(n)$ time,
    meaning that we can solve $k$-SUM in $n^{o(k)}+\Theta(n)$ time. For $k=\omega(1)$, this is $n^{o(k)}$ time, contradicting ETH.
\end{proof}

We note that Theorem~\ref{thrm:partition_hard} relies on the Exponential Time Hypothesis. Recently, OpenAI and Anthropic announced refutations of ETH by giving subexponential-time algorithms for 3SAT~\cite{OpenAIETH2026,AnthropicETH2026}. If either of these results is confirmed, our reduction still gives a hardness result under the assumption that $\mathrm{FPT}\neq\mathrm{W[1]}$. Indeed, $k$-SUM is $\mathrm{W[1]}$-hard when parameterized by $k$~\cite{AbboudLewiWilliams2014}, and our reduction runs in polynomial time and preserves $k$. Thus, $Pm \mid \mid C_{\max}$ with $k$-perturbed partition-based encoding is also $\mathrm{W[1]}$-hard when parameterized by $k$, even when $m=2$. Therefore, unless $\mathrm{FPT}=\mathrm{W[1]}$, the problem admits no fixed-parameter tractable algorithm with respect to $k$, that is, no algorithm running in $f(k) \cdot n^{O(1)}$ time for any computable function $f$.

A fundamental limitation of partition-based predictions seems to be that, although we know that at most $k$ jobs are erroneously assigned to each machine, we have no information to identify them. This raises the question of whether knowing exactly which jobs may be misplaced is sufficient to overcome the computational hardness of scheduling.

To investigate this, we assume that we have access to a predicted assignment of all jobs to machines that also identifies any job that may be wrongfully assigned, without having any knowledge of where these jobs should be assigned. If the number of machines, $m$, is constant and we assume that there are at most $k$ errors in each machine, then minimizing the makespan can be straightforwardly solved by enumerating at most $m^{mk}$ assignments\footnote{With at most $k$ errors on each of the $m$ machines, there is a total of $mk$ errors, each of which has $m$ possible placements.}, which is polynomial if $k = \bigO(\log n)$. However, we prove that even if each machine contains at most $3$ jobs that we know may be erroneously placed, then $P \mid \mid C_{\max}$ (where the number of machines is not constant) remains strongly \NP-complete.

\begin{definition}[$P \mid \mid C_{\max}$ with $k$-perturbed error-aware partition-based encoding]
    Let~$I$ be an instance of $P \mid \mid C_{\max}$ and $t$ be a target makespan. Let $S_1,\dots,S_m$ be a perturbed partition and $S_1^{err} \subseteq S_1,\dots,S_m^{err}\subseteq S_m$ be the potentially misplaced jobs for each machine, with $|S_i^{err}|\leq k$ for all $i \in [m]$. Is there an $m$-way partition of jobs to machines with makespan at most $t$?
\end{definition}

We get this hardness result via a reduction from $3$-Partition. Intuitively, knowing which jobs may be misplaced without having any information on which of the $m$ machines they belong to leaves each misplaced job with too many choices, as $m$ is part of the input.

\begin{restatable}{theorem}{thmthreepartition}
\label{theorem:3partition}
    $P \mid \mid C_{\max}$ with $k$-perturbed error-aware partition-based encoding is strongly \NP-complete, even when $k=3$.
\end{restatable}

\section{Order-based predictions: Scheduling problems determined by job partition}
\label{sec:outer_DP}
We now focus on order-based predictions. We consider scheduling problems for which an optimal solution can be efficiently computed from a partition of the jobs among the $m$ machines ($m$-way partitioning). Assuming that $m$ is part of the input, we present a DP algorithm for makespan-like objectives, where the goal is to minimize the maximum value attained over all machines.\footnote{The algorithm can also be adapted to maximization problems.}

Given a $k$-perturbed order-based encoding of a (hidden) optimal solution to the problem, we show how to compute an optimal schedule. We first define the problem formally.

\begin{definition}[$m$-way partitioning with $k$-perturbed order-based encoding]
Let $J=\{j_1,\ldots,j_n\}$ be a set of jobs, let $m$ be the number of machines, and let $\sigma$ be a perturbed order-based encoding of a (hidden) optimal solution such that $|\sigma^{-1}(i) - (\sigma^*)^{-1}(i)| \leq k$ for all $i \in [n]$, where $\sigma^*$ is the order-based encoding of this optimal solution.
For each machine $i \in [m]$, let $f_i$ be an efficiently computable
function that returns the objective value when a given subset of jobs is assigned to machine~$i$.
The goal is to find a partition $S_1 \cup \cdots \cup S_m = J$, with $S_i \cap S_j = \emptyset$ for all $i \neq j$, minimizing
$\max_{i \in [m]} \; f_i \bigl(S_i\bigr)$.
\end{definition}

This formulation captures several standard makespan-minimization objectives whose value is determined once the partition of jobs among the machines is fixed.
First, for $P \mid \mid C_{\max}$ we define 
$f_i(S_i) = \sum_{j \in S_i} p_j$,
for every machine $i \in [m]$. Thus the objective $\max_{i \in [m]} f_i(S_i)$ is exactly the makespan of the schedule.
For unrelated machines, where job $j$ has processing time $p_{ij}$ on machine $i$, we set $f_i(S_i) = \sum_{j \in S_i} p_{ij}$.
Related machines are obtained as the special case $p_{ij} = p_j/s_i$, where $s_i$ is the speed of machine $i$. 
We now give our dynamic programming algorithm.

\begin{theorem} \label{thrm:outer_dp}
    Given an instance of $m$-way partitioning with a $k$-perturbed order-based encoding of an optimal schedule, we can construct a DP that solves the problem in time $\bigO(m \cdot n^2 \cdot 2^{4k} \cdot F)$, where $F \ge n$ is an upper bound on the time required to evaluate any function $f_i$ on a given subset.
\end{theorem}

\begin{proof}
    Let $\mathcal{P}_{m'}$ be the set of all possible subsets of jobs that could be assigned to the first $m'$ machines in an optimal schedule consistent with the given $k$-perturbed order-based encoding. 
    To construct this set, consider a cut position $c \in \{0,1,\dots, n\}$, namely a cut after the first $c$ jobs in the (hidden) optimal order. In what follows, all index intervals are
    implicitly intersected with $[1,n]$. Since each job is displaced by at most $k$ positions, any candidate subset $P_{m'} \in \mathcal{P}_{m'}$ corresponding to this cut must contain every job $j_{\sigma(q)}$
    with $q \le c-k$ and contain no job $j_{\sigma(q)}$ with $q \ge c+k+1$. The only undecided jobs are those $j_{\sigma(q)}$ with $q \in [c-k+1,c+k]$, and we only keep choices satisfying $|P_{m'}|=c$.
    Hence, for a fixed cut position $c$, $P_{m'}$ is determined by choosing which jobs from a set of at most $2k$ jobs belong to $P_{m'}$, subject to the cardinality constraint. Therefore, for each $c$ there are at most $2^{2k}$ possible choices. Iterating over all possible cut positions, the set $\mathcal{P}_{m'}$ has size $\bigO(n \cdot 2^{2k})$. Moreover, if $P^*_{m'}$ is the set of jobs assigned to the first $m'$ machines in the hidden optimal schedule, then $P^*_{m'}\in\mathcal{P}_{m'}$ for every $m'\in[m]$.

    We now define the DP table. For each $m' \in [m]$ and each $P_{m'} \in \mathcal{P}_{m'}$, let $T[m', P_{m'}]$ denote the minimum objective value over all schedules of the jobs in $P_{m'}$ on the first $m'$ machines. A set $P_{m'} \in \mathcal{P}_{m'}$ is compatible with a set
    $P_{m'+1} \in \mathcal{P}_{m'+1}$ if $P_{m'} \subseteq P_{m'+1}$.
    Indeed, any jobs assigned to the first $m'$ machines must also be among the jobs assigned to the first $m'+1$ machines, while the jobs in $P_{m'+1} \setminus P_{m'}$ are precisely those assigned to machine $m'+1$. 
    
    For $m'=1$, we set
    $T[1, P_1] = f_1(P_1)$, for every $P_1 \in \mathcal{P}_1$.
    
    For $m' \ge 2$, we define
        $T[m',P_{m'}]=
        \min_{\substack{P_{m'-1} \in \mathcal{P}_{m'-1}\\ P_{m'-1} \subseteq P_{m'}}}
        \max \bigl\{T[m'-1,P_{m'-1}], f_{m'}(P_{m'} \setminus P_{m'-1})\bigr\}$.
    
    The optimal objective value is $T[m,J]$, since the full set of jobs must be assigned to the $m$ machines. Note that the recurrence allows $P_{m'-1} = P_{m'}$, corresponding to an empty assignment to machine $m'$. If empty machines are not admissible, one can restrict the minimization to $P_{m'-1} \subsetneq P_{m'}$.

    For each $m' \in [m]$, the set $\mathcal{P}_{m'}$ has size $\bigO(n \cdot 2^{2k})$. Hence, the DP has $\bigO(m \cdot n \cdot 2^{2k})$ states in total. For each state $T[m', P_{m'}]$, we consider all candidate subsets $P_{m'-1} \in \mathcal{P}_{m'-1}$, of which there are at most $\bigO(n \cdot 2^{2k})$, and for each compatible subset we evaluate $f_{m'}(P_{m'} \setminus P_{m'-1})$ in time at most $F$. Consequently, the overall running time is 
    $ \bigO(m \cdot n^2 \cdot 2^{4k} \cdot F)$.
\end{proof}

Although Theorem~\ref{thrm:outer_dp} is stated for objectives of the form $\max_{i \in [m]} f_i(S_i)$, the same DP applies more generally. It suffices that the objective value on the first $m'$ machines can be obtained by combining the objective value on the first $m'-1$ machines with the contribution of machine $m'$ through a monotone aggregation operator. In that case, the recurrence is given by replacing the maximum operator with this aggregation operator. For example, for objectives of the form $\sum_{i \in [m]} f_i(S_i)$, the same recurrence applies after replacing the maximum operator by addition.

This captures, for example, objectives such as $\sum C_j$ and $\sum w_j C_j$, whenever an optimal schedule on a single machine can be computed efficiently after the assigned set of jobs is fixed. 
For $P \mid \mid \sum C_j$, $f_i(S_i)$ is the sum of completion times obtained by scheduling the jobs of $S_i$ in shortest processing time (SPT) order. Similarly, for $P \mid \mid \sum w_j C_j$ with positive weights, $f_i(S_i)$ is the weighted sum of completion times obtained by ordering the jobs of $S_i$ according to Smith's rule~\cite{smith}, that is, in non-decreasing order of $p_j / w_j$. 

In the previous problems $P\mid\mid C_{\max}$, $P\mid\mid\sum C_j$ and $P\mid\mid\sum w_j C_j$, computing $f_i(S_i)$ is straightforward, since the corresponding single-machine subproblems are polynomial-time solvable once $S_i$ is fixed. 
However, our approach is also able to tackle some problems that are already $\NP$-hard on a single machine: to do so, we use the order-based prediction to compute $f_i(S_i)$ efficiently. 

We illustrate this phenomenon for scheduling problems with release times. For $P \mid r_j \mid C_{\max}$, the framework applies directly: once the set of jobs assigned to each machine is fixed, the optimal single-machine makespan can be computed in polynomial time by scheduling jobs as early as possible, keeping the machine busy whenever at least one job is available. For $P \mid r_j \mid \sum w_j C_j$, the induced single-machine problem $1 \mid r_j \mid \sum w_j C_j$ is \NP-hard in general~\cite{lenstra}. Nevertheless, using the order-based prediction, we solve this single-machine subproblem in time $n^2\cdot 2^{\bigO(k)}$ and combine it with the DP of Theorem~\ref{thrm:outer_dp} to obtain an algorithm for $P \mid r_j \mid \sum w_j C_j$ running in time $m\cdot n^4\cdot 2^{\bigO(k)}$. The details are deferred to Appendix~\ref{app:release_times}.

\section[Order-based predictions: Scheduling with precedence constraints]{\textls[-6]{Order-based predictions: Scheduling with precedence constraints}}
\label{sec:order_prec}
In this section, we address scheduling problems where dependencies between different machines fundamentally change our ability to schedule a job at each position on its assigned machine. 
This is distinct from the problems covered by our algorithms in Section~\ref{sec:outer_DP}. There, once a partition of jobs to machines has been fixed, the problem can be solved independently on each machine. Here, we tackle settings in which scheduling a job at a specific time on one machine may directly affect when a job can be scheduled in another machine. An example where such dependencies arise is when precedence constraints are involved. Specifically, we study this within the $Pm \mid \mathrm{prec} \mid C_{\max}$ framework and later generalize our approach to problems with similar structure.

We again assume that we are provided with a $k$-perturbed order-based encoding of a (hidden) optimal solution to the problem and show how to compute an optimal schedule in pseudopolynomial time.\footnote{Recall that $Pm \mid \mathrm{prec} \mid C_{\max}$ is strongly \NP-hard, even when $m=2$ and all precedence constraints consist of chains~\cite{prec_strong_NP-hard}, meaning that such a result is not possible without predictions.} Before presenting our algorithm, we first formally define the problem. Although we focus on parallel identical machines, notice that we do not require the processing time of a job to be the same across different machines. This means that our results can be straightforwardly generalized to address settings with unrelated machines.

\begin{definition}[$Pm \mid \mathrm{prec} \mid C_{\max}$ with $k$-perturbed order-based encoding]
    Let $I$ be an instance of $Pm \mid \mathrm{prec} \mid C_{\max}$ and $\sigma$ be a perturbed order-based encoding of a (hidden) optimal solution for $I$ such that $|\sigma^{-1}(i) - (\sigma^*)^{-1}(i)| \leq k$ for all $i \in [n]$, where $\sigma^*$ is the order-based encoding of said optimal solution. The goal is to find a schedule that minimizes the makespan.
\end{definition}

\begin{restatable}{theorem}{thmorderprecdp}
\label{thrm:order_prec_dp}
Given an instance of $Pm \mid \mathrm{prec} \mid C_{\max}$ with a $k$-perturbed order-based encoding, where $k \ge 1$, we can retrieve in time 
$\bigO\Bigl(\left(n\cdot 2^{2k}\right)^{2m} \cdot \bigl(k \cdot \sum_{i=1}^n p_i\bigr)^{m+1} \Bigr)$ a makespan minimizing schedule .
\end{restatable}

\begin{proof}[Proof sketch] The full proof is deferred to Appendix~\ref{app:thm14}.
Our algorithm consists of three distinct layers: an outer exhaustive search that partitions the jobs to machines (using an approach similar to that of Theorem~\ref{thrm:outer_dp}), a reduction from the minimization problem to a decision one (i.e., if a specific makespan is achievable by a feasible and consistent schedule), and a recursive inner DP algorithm that finds the optimal ordering of jobs inside each machine, adding idle time when needed.

\paragraph{The inner DP.} Given a partition $(S_1,\dots,S_m)$ of jobs to machines, the inner DP recursively determines the ordering of jobs and the idle time inside each machine. 
For each machine $m' \in [m]$, we consider its states as all triplets $(S_{m'}',j_{m'}^\star,t_{m'})$, where $S_{m'}' \subseteq S_{m'}$ is the (ordered) subset of jobs that will be scheduled first on this machine, $j_{m'}^\star \in S_{m'}'$ is the job from this subset that will be scheduled last, and $t_{m'}$ is the target completion time for this machine. 

The DP table is indexed by an $m$-tuple $((S_{1}',j_{1}^\star,t_{1}),\dots,(S_{m}',j_{m}^\star,t_{m}))$ of these machine states. A cell stores \texttt{yes} if it is feasible to schedule (in a way consistent with the $k$-perturbed order-based encoding~$\sigma$) every job subset $S_{m'}'$ on machine $m'$ so that job $j_{m'}^\star$ is scheduled last and the machine finishes exactly at time $t_{m'}$, for all $m' \in [m]$. Otherwise, it holds \texttt{no}. 

The initial state of the DP table is one where no jobs have been scheduled on any machine and all machines must complete at time~$0$. The cell for this state holds \texttt{yes} and is represented by the $m$-tuple where $S_{m'}' = \emptyset$, no job is scheduled last and $t_{m'}=0$ for all $m' \in [m]$. Conversely, any cell where there exists a machine $m' \in [m]$ with $S_{m'}'=\emptyset$ but $t_{m'}>0$ holds \texttt{no}.

During the main recursive phase, the algorithm sequentially considers all $m$-tuples $((S_{1}',j_{1}^\star,t_{1}),\allowbreak \dots,(S_{m}',j_{m}^\star,t_{m}))$. For each of them, it selects the machine $\ell \in [m]$ whose currently fixed last job, $j_\ell^\star$, has the latest starting time, meaning that
$
\ell = \argmax_{m' \in [m]} \{t_{m'} - p_{m'}^\star\},
$
where $p_{m'}^\star$ denotes the processing time of $j_{m'}^\star$, for $m' \in [m]$. We break ties arbitrarily.

If there exists an $m' \in [m]$ with $t_{m'} > t_\ell - p^{\star}_\ell$ such that $j^{\star}_{m'}$ must precede $j^{\star}_\ell$ or some job in $\left(\bigcup_{i=1}^m S_i'\right)\setminus\{j_\ell^\star\}$ must succeed $j_\ell^\star$, the cell holds \texttt{no} for this $m$-tuple, as a precedence constraint is violated. Otherwise, job $j_\ell^\star$ is safely scheduled to begin at time $t_\ell - p^{\star}_\ell$ on machine $\ell$. This assignment, together with the recursive search on machine $\ell$, guarantees that no precedence constraints involving $j_\ell^\star$ are violated. Every non-last job on another machine finishes by the starting time of that machine's last job, which is at most $t_\ell-p_\ell^\star$, while the conditions above exclude remaining successors of $j_\ell^\star$ and last jobs that finish too late to precede it. To determine the value the DP cell must hold for this $m$-tuple, we examine already computed cells corresponding to states that do not include $j_\ell^\star$. Specifically, we search for all DP states identical to the current one, except for the triplet of machine~$\ell$. The state for machine $\ell$ is of the form $(S_\ell' \setminus \{j_\ell^\star\}, j_\ell^{\star'},t_\ell')$, where $j_\ell^{\star'}$ is the last job of the subset $S_\ell' \setminus \{j_\ell^\star\}$ and $t_\ell'$ is the target completion time for this subset. Since $\sigma$ is a $k$-perturbed order-based encoding, there are at most $\bigO(k)$ candidates for $j_\ell^{\star'}$. Additionally, to account for all feasible idle times, we consider all possible values in the range $[\sum_{j_i \in S_\ell' \setminus \{j_\ell^\star\}} p_i,\ t_\ell - p_\ell^\star]$ for $t_\ell'$. This lookup process takes $\bigO(k \cdot \sum_{i=1}^n p_i)$ time. The DP cell holds \texttt{yes} for the current $m$-tuple if and only if at least one of the searched cells holds \texttt{yes}. \label{thrm:inner_dp}
\end{proof}

In the algorithm of Theorem~\ref{thrm:order_prec_dp}, each DP cell holds only \texttt{yes} or \texttt{no} to indicate feasibility. By instead storing the partial objective value in each cell, we can directly generalize this approach to tackle more sophisticated objective functions (or even problems), such as $\sum w_j C_j$ (see Appendix~\ref{app:sumwjcj}).

\section{Limitations of order-based predictions}
\label{sec:limitations_order}

A $k$-perturbed order-based encoding has proven to be more informative compared to perturbed orders in the other two encodings, since it appears to be more resistant to noise. Although it is powerful, the order-based prediction is not a silver bullet for all scheduling problems. Specifically, we prove that it cannot assist in efficiently solving $P \mid \mathrm{prec} \mid C_{\max}$ when the number of machines is part of the input, unless $\P = \NP$, via a reduction from $3$-SAT. This reduction holds even when we are given a $2$-perturbed order-based encoding of an optimal schedule. The proof is deferred to Appendix~\ref{app:limiations_order}.

\begin{restatable}{theorem}{thmhardorderbased}
\label{thrm:hard_order-based}
    $P \mid \mathrm{prec} \mid C_{\max}$ with $2$-perturbed order-based encoding is strongly $\NP$-complete.
\end{restatable}

\section{Conclusion}
\label{sec:conclusion}

In this work, we analyzed which noisy predictions enable efficient exact schedule recovery. While greedy and partition-based models are highly noise-sensitive, order-based predictions allow polynomial-time recovery for $k=\bigO(\log n)$ and pseudopolynomial-time recovery for fixed machines with precedence constraints. However, hardness results for $P\mid \mathrm{prec}\mid C_{\max}$ highlight inherent limitations even for order-based information.

An interesting direction for future work is to understand how far the order-based approach can be pushed beyond scheduling. The framework of Bampis et al.~\cite{icml_paper} applies to permutation optimization problems, including single-machine scheduling. Our partitioning ideas allow this approach to extend to problems whose solutions consist of several ordered structures rather than a single ordering, as long as the contribution of each structure can be evaluated independently once the partition is fixed. This includes, for example, multi-route variants of Hamiltonian path or TSP problems. A natural question is to identify the broader range of problems to which this approach applies.

\section*{Acknowledgments and Disclosure of Funding}
    Dimitris Fotakis is partially supported by the ERC Consolidator Grant COMFORMAL (No. 101232091). Michalis Xefteris is supported by a Humboldt Research Fellowship
    for Postdoctoral Researchers from the Alexander von Humboldt Foundation.

\section*{Disclaimer about LLM usage} LLMs were used only for writing and editing assistance, and not as part of the core methodology or results of the paper.

\bibliographystyle{plainnat}
\bibliography{bibliography}

\appendix

\section{Deferred proofs for Section~\ref{sec:hardness}}
\label{app:hardness}

\subsection{Proof of Theorem~\ref{thrm:greedy_hard}}

\thmgreedyhard*

\begin{proof}
    The problem is in \NP, since any proposed schedule can be straightforwardly checked for feasibility. We now reduce $P2 \mid \mid C_{\max}$ to $P2 \mid \mid C_{\max}$ with a $1$-perturbed greedy order.

    Given any set of jobs $J = \{j_1,...,j_n\}$ with their respective processing times $p_1,\dots,p_n$ that constitutes a non-trivial $P2 \mid \mid C_{\max}$ instance, we will construct a set $J'$ of $2n$ jobs to be assigned to two machines, as well as a permutation $\sigma$ of these jobs. We will then prove that there exists an optimal greedy ordering $\sigma^*$ with $|\sigma^{-1}(i)-(\sigma^*)^{-1}(i)|\leq 1$ for all $i \in [2n]$, i.e. an ordering such that, if the jobs of $J'$ are scheduled greedily in the order of $\sigma^*$, the resulting schedule achieves the minimum possible makespan. Lastly, we will prove that a target makespan $t$ is achieved for $J$ if and only if a makespan of $t'$ is achieved for $J'$.

    Let $B=\sum_{i=1}^n p_i+1$. We define the set of jobs $J'=\big\{j_i' \mid i \in [2n]\big\}$ with processing times $\{p_1+B, B, p_2+B,..., p_n+B, B\}$. For each $i \in [n]$, we denote by $c_i$ the pair of jobs $\{j_{2i-1}',j_{2i}'\}$, with processing times $p_i+B$ and $B$, respectively. Let $\sigma = (j_1',j_2',\dots,j_{2n}')$ refer to the order in which the jobs are listed in the above definition. We prove that there exists an optimal greedy ordering $\sigma^*$ for the jobs of $J'$ such that $|\sigma^{-1}(i)-(\sigma^*)^{-1}(i)|\leq 1$, for all $i \in [2n]$.

    Let $S_1^*$ and $S_2^*$ be the makespan-minimizing subsets that partition $J$, resulting in a makespan of $C^*$. It follows directly that the minimum makespan allocation for $J'$ schedules exactly $n$ jobs to each machine and cannot obtain a makespan lower than $C^*+n B$. Indeed, since every job has processing time at least $B$, a schedule assigning $n+1$ jobs to one machine would result in a makespan of at least $(n+1)B$. However, scheduling exactly $n$ jobs to each machine results in a makespan of at most $nB +B-1$. Therefore, if we could obtain a schedule allocating $n$ jobs to each machine with makespan $C^*+nB$, then by subtracting $B$ from the processing time of each job we would get a set of jobs equivalent to $J$ with makespan $C^*$, thus optimally solving $P2 \mid \mid C_{\max}$. Hence, minimizing the makespan of $J'$ is equivalent to scheduling the jobs in a way such that the first machine contains all the jobs with processing times $P_1^* = \{B+p_i \mid j_i \in S_1^*\}$ and the second contains all the jobs $P_2^* = \{B+p_i \mid j_i \in S_2^*\}$. Then, assigning $|S_2^*|$ of the jobs with processing time $B$ to the first machine and $|S_1^*|$ of them to the second one will yield the target makespan of $C^*+nB$. 

    It remains to show that there exists a permutation $\sigma^*$ with $|\sigma^{-1}(i)-(\sigma^*)^{-1}(i)|\leq 1$ for all $i \in [2n]$ that forces the greedy algorithm to place all the jobs of $P_1^*$ in the first machine and all the jobs of $P_2^*$ in the second one. We do that by induction on the number of pairs $c_i = \{p_i+B,B\}$ processed. Since the two machines are identical, we assume without loss of generality that the tie breaking rule is to always schedule the job on the first machine.

    \emph{Basis $(x=1)$}:In the beginning, the tie breaking rule applies. Thus, if $p_1+B \in P_1^*$, then job $j_1'$ will be scheduled in the first machine and job $j_2'$ will be scheduled on the second machine. If $p_1+B \in P_2^*$, then swapping jobs $j_1'$ and $j_2'$ will place job $j_1'$ on the second machine, by the same argument.
    
    \emph{Inductive Hypothesis}: Suppose that $\sigma^*$ forces the greedy algorithm to correctly schedule all couples up to $c_{x-1}$, under the condition $|\sigma^{-1}(i)-(\sigma^*)^{-1}(i)|\leq 1$, for every $i \in [2(x-1)]$.
    
    \emph{Inductive Step}: We prove that the algorithm is forced to correctly schedule $c_x$, if $\sigma^*$ can either maintain or swap the relative order of jobs in $c_x$. Recall that $c_x$ contains the pair of jobs $\{j_{2x-1}',j_{2x}'\}$, with processing times $p_x+B$ and $B$, respectively. Without loss of generality, assume that $j_x \in S_1^*$, since the argument for the other case would be analogous. We differentiate between two cases:

    If machine~$1$ becomes available first, then the greedy algorithm will schedule job $j_{2x-1}'$ to it. Since the first machine now contains one more job than the second one, it is certain to be available last, by construction of $B$. Thus, the algorithm will schedule job $j_{2x}'$ to the second machine.

    If machine~$2$ becomes available first, then the greedy algorithm will incorrectly try to schedule job $j_{2x-1}'$ to it. Thus, we need to swap jobs $j_{2x-1}'$ and $j_{2x}'$. After swapping, the algorithm places job $j_{2x}'$ to machine $2$, making it the one that becomes available last. Hence, when the greedy algorithm processes job $j_{2x-1}'$, it schedules it on the first machine. 
\end{proof}

\subsection{Proof of Theorem~\ref{theorem:3partition}}

\thmthreepartition*

\begin{proof}
    The problem is in \NP, since any proposed schedule can be directly checked for feasibility. We now reduce $3$-Partition to $P \mid \mid C_{\max}$ with a $3$-perturbed error-aware partition-based encoding.

    Suppose that we are given a (multi)set $A = \{a_1,\dots,a_n\}$ of positive integers, where $n = 3m$. Let $B = \sum_{i=1}^n a_i$ and assume that $B/(4m) <a_i < B/(2m)$, for all $i \in [n]$. The goal is to partition $A$ into $m$ disjoint triplets $S_1,\dots,S_m$ such that every triplet has a sum of $B/m$. We transform this into an instance~$I$ of $P \mid \mid C_{\max}$ with $3$-perturbed error-aware partition-based encoding, setting the target makespan to $t = B/m$. Each $a_i \in A$ corresponds to a job $j_i \in J$ with processing time $p_i = a_i$, for all $i \in [n]$. For the predicted assignment, we arbitrarily assign three jobs to each machine, setting $S_{m'} = \{j_{3m'-2},j_{3m'-1},j_{3m'}\}$, for all $m' \in [m]$. Since each machine is assigned exactly three jobs, at most three jobs may be misplaced. Therefore, we assume that $S_{m'}^{err}=S_{m'}$ with $|S_{m'}^{err}|=3$, for all $m' \in [m]$, and the error-aware constraint is always trivially satisfied.

    We prove that $A$ is a yes-instance for $3$-Partition if and only if a makespan of $B/m$ is achievable for~$I$. First, assume that $A$ is a yes-instance, meaning that there exists a partition of $A$ into $m$ triplets, with each summing to $B/m$. Assigning the jobs corresponding to the elements of each triplet to distinct machines admits a schedule where every machine finishes exactly at time $B/m$, thus achieving the target makespan and making~$I$ a yes-instance. Conversely, assume that a makespan of $t=B/m$ is achievable for~$I$. Since the total processing time of all jobs is $B$, every machine must have a load of exactly $B/m$ to achieve a makespan of $t$. Notice that $B/(4m) < p_i < B/(2m)$, for all $i \in [n]$, hence for that to be possible each machine must be assigned exactly three jobs. Therefore, forming triplets with the elements of $A$ that correspond to the jobs assigned to each machine admits a partition of $A$, with each triplet having a sum of $B/m$, thus making $A$ a yes-instance for $3$-Partition.
\end{proof}

\section{Missing proofs for Section~\ref{sec:outer_DP}}
\label{app:release_times}

\subsection{Scheduling with release times}
 
First, we observe that makespan minimization with release times fits directly into the framework developed in Section~\ref{sec:outer_DP}. Indeed, for $P \mid r_j \mid C_{\max}$, once the set $S_i$ of jobs assigned to machine $i$ is fixed, the optimal single-machine makespan on that machine can be computed in polynomial time. If $S_i = \emptyset$, we set $f_i(S_i) = 0$. Otherwise, it suffices to schedule the jobs of $S_i$ as early as possible, keeping the machine busy whenever at least one job is available. For example, one may schedule the jobs in non-decreasing order of release times. Then $f_i(S_i)$ can be defined as the completion time of this single-machine schedule, and Theorem~\ref{thrm:outer_dp} applies.

We now turn to the more challenging objective $P \mid r_j \mid \sum w_j C_j$, with positive weights. Unlike the makespan case, once the partition of jobs among the machines is fixed, the optimal ordering on each machine is not immediate. Instead, for each machine, the induced subproblem is a single-machine problem of the form 
$1 \mid r_j \mid \sum w_j C_j$.

The single-machine problem $1 \mid r_j \mid \sum w_j C_j$ is \NP-hard in general~\cite{lenstra}, even without weights.
Nevertheless, we show that the problem can be solved efficiently when a perturbed order-based encoding is available.

\begin{lemma} \label{lem:single_release_time}
    Given an instance of $1 \mid r_j \mid \sum w_j C_j$ with a $k$-perturbed order-based encoding of an optimal schedule, the problem can be solved in time $n^2 \cdot 2^{\bigO(k)}$.
    
\end{lemma}

\begin{proof}
    Let $\sigma$ be the given $k$-perturbed order-based encoding. We construct the schedule position by position, where positions refer to the order of the jobs.

    For every $i\in \{0, \dots, n\}$, let $\mathcal{R}_{i}$ be the set of all possible subsets of jobs that could be scheduled in the first $i$ positions of a schedule consistent with the given $k$-perturbed order-based encoding. In what follows, all index intervals are implicitly intersected with $[1,n]$. Since each job is displaced by at most $k$ positions, any candidate subset $R_{i} \in \mathcal{R}_{i}$ must satisfy the following:

    \begin{itemize}
        \item each job $j_{\sigma(q)}$ with $q \le i-k$ must belong to $R_i$,

        \item no job $j_{\sigma(q)}$ with $q \ge i+k+1$ can belong to $R_i$,

        \item the jobs $j_{\sigma(q)}$ with $q \in [i-k+1,i+k]$ may or may not belong to $R_i$
    \end{itemize}
    and, moreover, $|R_i|=i$.

    Hence, for a fixed position $i$, $R_i$ is determined by choosing which jobs from a set of at most $2k$ jobs belong to $R_i$, subject to the cardinality constraint $|R_i|=i$. Therefore, for each $i$ there are at most $2^{2k}$ possible choices. Iterating over all possible positions, the total number of such states is $\bigO(n \cdot 2^{2k})$.

    We now define the DP table. For each $i \in \{0, \dots, n\}$ and each $R_i \in \mathcal{R}_i$, we maintain a Pareto list $L[i,R_i]$ of pairs $(t,C)$,
    where $t$ is the completion time of the last scheduled job and $C$ is the accumulated weighted sum of completion times of the jobs in $R_i$. A pair $(t_1, C_1)$ dominates another pair $(t_2, C_2)$ if 
    $$
        t_1 \le t_2 
        \qquad \text{and} \qquad
        C_1 \le C_2.
    $$

    Dominated pairs can be discarded. From the same state $R_i$, the set of remaining jobs is fixed. Therefore, if one partial schedule has both no larger current completion time and no larger accumulated objective value, then it can only lead to a solution that is at least as good as the other partial schedule.

    For $i=0$, we set
    $L[0, \emptyset] = \{ (0,0)\}$.
    A set $R_i \in \mathcal{R}_i$ is compatible with a set $R_{i+1} \in \mathcal{R}_{i+1}$ if $R_i \subseteq R_{i+1}$. In this case, the unique job in $R_{i+1} \setminus R_i$ is the job scheduled in position $i+1$. Let this job be $u$. For every pair $(t_{\mathrm{prev}}, C_{\mathrm{prev}}) \in L[i, R_i]$, we create the pair
    $$
        t_{\mathrm{new}} = \max\{t_{\mathrm{prev}}, r_u\} + p_u,
    $$
    and
    $$
        C_{\mathrm{new}} = C_{\mathrm{prev}} + w_u \cdot t_{\mathrm{new}}.
    $$
    We insert $(t_{\mathrm{new}}, C_{\mathrm{new}})$ into $L[i+1, R_{i+1}]$ and remove dominated pairs. 

    By construction, the compatible chains $R_0 \subseteq R_1 \subseteq \cdots \subseteq R_n$ are exactly the job orders that are $k$-perturbed with respect to $\sigma$.

    It remains to bound the size of each Pareto list. Fix a state $R_i \in \mathcal{R}_i$. Consider any partial schedule reaching this state, and let $b \le i$ be the first position of its last busy period. Let $v$ be the first job of this busy period, and let $B$ be the set of jobs processed during this busy period. Then
    $$
        t = r_v + \sum_{j \in B} p_j.
    $$
    Fix the position $b$. The jobs processed before this last busy period are precisely the jobs scheduled in positions $1,\dots, b-1$. Therefore, they form some set $R_{b-1} \in \mathcal{R}_{b-1}$, and the set of jobs in the last busy period is 
    $B = R_i \setminus R_{b-1}$.
    For the fixed value of $b$, there are at most $|\mathcal{R}_{b-1}| \le 2^{2k}$ possible choices for $R_{b-1}$, and thus at most $2^{2k}$ possible choices for $B$. Moreover, the first job $v$ of the last busy period is scheduled in position $b$, and therefore its predicted position belongs to $[b-k, b+k]$, giving at most $2k+1$ choices for $v$.

    Thus, for a fixed $b$, there are at most $(2k+1)2^{2k}$ possible values of $t$. Since there are at most $n$ choices for $b$, the number of possible completion times in a fixed state is at most 
    $n(2k+1)2^{2k} = n \cdot 2^{\bigO(k)}$.
    For each completion time $t$, we keep only the smallest accumulated value $C$. Consequently, each Pareto list has size at most $n \cdot 2^{\bigO(k)}$.

    There are $\bigO(n\cdot 2^{2k})$ states. For each state, we consider at most $\bigO(2^{2k})$ compatible successor states, and each transition processes a Pareto list of size at most $n \cdot 2^{\bigO(k)}$. Therefore, the total running time is $n^2 \cdot 2^{\bigO(k)}$.
    The optimum value is obtained by taking the minimum value of $C$ over all pairs in the Pareto list of the terminal state $R_n=J$. Since the predicted order is $k$-perturbed with respect to an optimal schedule, this optimal schedule is among the schedules considered by the DP. Consequently, the DP returns an optimal solution.
\end{proof}

We now use this single-machine algorithm as a subroutine in the outer DP.

\begin{theorem}
\label{thrm:multiple_release}
    Given an instance of $P \mid r_j \mid \sum w_j C_j$ with positive weights and a $k$ perturbed order-based encoding of an optimal schedule, the problem can be solved in time $m\cdot n^4\cdot 2^{\bigO(k)}$.
\end{theorem}

\begin{proof}
    Let $\sigma$ be the given $k$-perturbed order-based encoding. We use the outer DP of Theorem~\ref{thrm:outer_dp}. For a transition from $P_{m'-1}\in\mathcal{P}_{m'-1}$ to $P_{m'}\in\mathcal{P}_{m'}$, the jobs assigned to machine $m'$ are
    $S_{m'}=P_{m'}\setminus P_{m'-1}$.
    To evaluate this transition, we restrict $\sigma$ to the jobs of $S_{m'}$ and run the single-machine algorithm of Lemma~\ref{lem:single_release_time} with parameter $2k$. This gives the best schedule of $S_{m'}$ among those whose order is $2k$-perturbed with respect to the restricted order.

    It remains to justify that this is sufficient for the true (hidden) optimal partition. Let $\sigma^*$ be the hidden optimal order. If $\sigma$ is $k$-perturbed with respect to $\sigma^*$, then, after restricting both orders to the same subset of jobs, the two restricted orders are $2k$-perturbed.

    To see this, fix a job $j_x$ that remains after the restriction. The position of $j_x$ in a restricted order depends only on the jobs that remain before it. Jobs that are on the same side of $j_x$ in both $\sigma$ and $\sigma^*$ affect its position in the same way in both restricted orders. Thus, only jobs that cross $j_x$ can change its displacement after restriction.

    Consider first the jobs that appear before $j_x$ in $\sigma$ but after $j_x$ in $\sigma^*$. Any such job must be among the $2k$ jobs immediately before $j_x$ in $\sigma$. Otherwise, even if this job moves $k$ positions to the right and $j_x$ moves $k$ positions to the left, it cannot pass $j_x$. Hence there are at most $2k$ such jobs. Symmetrically, there are at most $2k$ jobs that appear after $j_x$ in $\sigma$ but before $j_x$ in $\sigma^*$.

    After restricting to a subset of jobs, the displacement of $j_x$ is the difference between the numbers of remaining crossing jobs of these two types. Both numbers are between $0$ and $2k$, so this difference has absolute value at most $2k$.

    Thus, for every transition corresponding to the true optimal partition, the single-machine algorithm considers the restriction of the true optimal order to the corresponding machine. Hence the outer DP can recover the value of the global optimum. Every schedule considered by the DP is feasible, so the DP cannot return a value smaller than the optimum.

    The outer DP has $\bigO(m\cdot n^2\cdot 2^{4k})$ transitions. Each transition calls Lemma~\ref{lem:single_release_time} with parameter $2k$, and therefore takes time $n^2 \cdot 2^{\bigO(k)}$. Consequently, the total running time is
    $
        m\cdot n^4\cdot 2^{\bigO(k)}.
    $
\end{proof}

\section{Missing proof of Section~\ref{sec:order_prec}}

\subsection{Proof of Theorem~\ref{thrm:order_prec_dp}}\label{app:thm14}

\thmorderprecdp*

\begin{proof}
Our algorithm consists of three distinct layers: an outer exhaustive search that partitions the jobs to machines, a reduction from the minimization problem to a decision one, and an inner DP algorithm that finds the optimal ordering of jobs inside each machine, adding idle time when needed. Note that, throughout this proof, we use $S_{m'}$, with $m' \in [m]$, to denote both a set and an ordered set of jobs. This allows us to seamlessly apply standard set operations (e.g. subset of, set difference and union) directly to ordered sets, without over-complicating the notation.

\paragraph{The exhaustive search.} We utilize an approach similar to that of Theorem~\ref{thrm:outer_dp} to find all consistent partitions of jobs. We again define $\mathcal{P}_{m'}$ as the set of all possible (ordered) subsets of jobs that could be assigned to the first $m'$ machines in an optimal schedule consistent with the given $k$-perturbed encoding. 
For each $S_1 \in \mathcal{P}_{1}$, representing an (ordered) set of jobs assigned to the first machine, the exhaustive search iteratively constructs all valid machine assignments by fixing 
\[
S_{m'} = P_{m'} \setminus \bigcup_{i=1}^{m'-1}S_{i} \textnormal{\qquad s.t. \qquad} P_{m'} \in \mathcal{P}_{m'},\ \bigcup_{i=1}^{m'-1}S_{i} \subseteq P_{m'},\ m' \in [2,m].
\]

Recall that $\mathcal{P}_{m'}$ has a size of $\mathcal{O}(n \cdot 2^{2k})$. Since each $S_{m'}$ is derived by removing elements from some $P_{m'} \in \mathcal{P}_{m'}$, the number of feasible candidate assignments to any machine shares the same bound. Therefore, iterating through all feasible (and consistent with the given $k$-perturbed encoding) ways of assigning jobs to machines takes $\bigO\left( \left(n \cdot 2^{2k}\right)^m \right)$ time.

\paragraph{Reducing to a decision problem.} For each feasible partition $(S_1,\dots,S_m)$ of jobs to machines found by the outer exhaustive search, where each $S_{m'}$ represents a $k$-perturbed ordering of jobs inside machine~$m' \in [m]$, we want to minimize the makespan. We reduce this to a decision problem, asking whether there exists a feasible schedule such that machine~$m'$ completes its processing at exactly time~$t_{m'}$, where $t_{m'}$, for $m' \in [m]$, are non-negative integers. We solve this decision problem using the following inner DP. Once the DP table is filled, it suffices to search its last level (representing states where all jobs have been scheduled) to find the cell storing \texttt{yes} that admits the minimum makespan. Since this search requires less time than computing the DP table itself, it does not affect the asymptotic complexity.

\paragraph{The inner DP.} Given a partition $(S_1,\dots,S_m)$ of jobs to machines, the inner DP recursively determines the ordering of jobs and the idle time inside each machine. 
For each machine $m' \in [m]$, we consider its states as all triplets $(S_{m'}',j_{m'}^\star,t_{m'})$, where $S_{m'}' \subseteq S_{m'}$ is the (ordered) subset of jobs that will be scheduled first on this machine, $j_{m'}^\star \in S_{m'}'$ is the job from this subset that will be scheduled last, and $t_{m'}$ is the target completion time for this machine. 

The DP table is indexed by an $m$-tuple $((S_{1}',j_{1}^\star,t_{1}),\dots,(S_{m}',j_{m}^\star,t_{m}))$ of these machine states. A cell stores \texttt{yes} if it is feasible to schedule (in a way consistent with the $k$-perturbed order-based encoding~$\sigma$) every job subset $S_{m'}'$ on machine $m'$ so that job $j_{m'}^\star$ is scheduled last and the machine finishes exactly at time $t_{m'}$, for all $m' \in [m]$. Otherwise, it holds \texttt{no}. 

The initial state of the DP table is one where no jobs have been scheduled on any machine and all machines must complete at time~$0$. The cell for this state holds \texttt{yes} and is represented by the $m$-tuple where $S_{m'}' = \emptyset$, no job is scheduled last and $t_{m'}=0$ for all $m' \in [m]$. Conversely, any cell where there exists a machine $m' \in [m]$ with $S_{m'}'=\emptyset$ but $t_{m'}>0$ holds \texttt{no}.

During the main recursive phase, the algorithm sequentially considers all $m$-tuples $((S_{1}',j_{1}^\star,t_{1}),\allowbreak \dots,(S_{m}',j_{m}^\star,t_{m}))$. For each of them, it selects the machine $\ell \in [m]$ whose currently fixed last job, $j_\ell^\star$, has the latest starting time, meaning that
\[
\ell = \argmax_{m' \in [m]} \{t_{m'} - p_{m'}^\star\},
\]
where $p_{m'}^\star$ denotes the processing time of $j_{m'}^\star$, for $m' \in [m]$. We break ties arbitrarily. Two distinct cases arise:

\subparagraph{Infeasibility.} If there exists an $m' \in [m]$ with $t_{m'} > t_\ell - p^{\star}_\ell$ such that $j^{\star}_{m'}$ must precede $j^{\star}_\ell$, then a precedence constraint is violated. Indeed, since job $j^{\star}_{m'}$ on machine $m'$ is required to finish at $t_{m'}$, it is impossible for $j^{\star}_\ell$ to start at $t_\ell - p^{\star}_\ell$. Precedence constraints are also violated if some job in $\left(\bigcup_{i=1}^m S_i'\right)\setminus\{j_\ell^\star\}$
must succeed $j_\ell^\star$. In both cases, the cell holds \texttt{no} for this $m$-tuple.

\subparagraph{Recursive search.} Otherwise, job $j_\ell^\star$ is safely scheduled to begin at time $t_\ell - p^{\star}_\ell$ on machine $\ell$. This assignment, together with the recursive search on machine $\ell$, guarantees that no precedence constraints involving $j_\ell^\star$ are violated. Every non-last job on another machine finishes by the starting time of that machine's last job, which is at most $t_\ell-p_\ell^\star$, while the conditions above exclude remaining successors of $j_\ell^\star$ and last jobs that finish too late to precede it. To determine the value the DP cell must hold for this $m$-tuple, we examine already computed cells corresponding to states that do not include $j_\ell^\star$. Specifically, we search for all DP states identical to the current one, except for the triplet of machine~$\ell$. The state for machine $\ell$ is of the form $(S_\ell' \setminus \{j_\ell^\star\}, j_\ell^{\star'},t_\ell')$, where $j_\ell^{\star'}$ is the last job of the subset $S_\ell' \setminus \{j_\ell^\star\}$ and $t_\ell'$ is the target completion time for this subset. Since $\sigma$ is a $k$-perturbed order-based encoding, there are at most $\bigO(k)$ candidates for $j_\ell^{\star'}$. Additionally, to account for all feasible idle times, we consider all possible values in the range $[\sum_{j_i \in S_\ell' \setminus \{j_\ell^\star\}} p_i,\ t_\ell - p_\ell^\star]$ for $t_\ell'$. This lookup process takes $\bigO(k \cdot \sum_{i=1}^n p_i)$ time. The DP cell holds \texttt{yes} for the current $m$-tuple if and only if at least one of the searched cells holds \texttt{yes}.

We now calculate the time complexity of this procedure. Since the search to determine the value of each cell takes $\bigO(k \cdot \sum_{i=1}^n p_i)$ time, computing the entire table requires multiplying the total number of states by this cost. Recall that the DP table is indexed by an $m$-tuple of triplets, each one of which is of the form $(S_{m'}', j_{m'}^\star, t_{m'})$ with $m' \in [m]$. We bound the number of possible subsets $S_{m'}'$ as follows: suppose that $S_{m'}'$ has a cardinality of $|S_{m'}'| = c \leq n$. We seek to find the number of different subsets of cardinality $c$ such that $S_{m'}' \subseteq S_{m'}$, given that $S_{m'}$ is a $k$-perturbed ordering. Since no job in $\sigma$ is displaced by more than $k$ positions, the first $c-k$ jobs of $S_{m'}$ are always included in $S_{m'}'$. However, jobs in positions $[c-k+1,\ c+k]$ of $S_{m'}$ may or may not belong in $S_{m'}'$. For $|S_{m'}'| = c$ to hold, the algorithm must select exactly $\min\{c,k\}$ jobs from a set of at most $2k$ jobs, resulting in $\binom{2k}{k} \leq 2^{2k}$ unique such (ordered) subsets of jobs for each cardinality $c \leq n$. Therefore, for a given assignment of jobs to a machine $m' \in [m]$, there are $\bigO(n \cdot 2^{2k})$ feasible and consistent job subsets. Within each such subset, at most $\bigO(k)$ jobs can be consistently scheduled last. Furthermore, the inner DP iteratively considers all integers from $0$ to $\sum_{i=1}^n p_i$ as target completion times, meaning that the number of possible targets is $\bigO(\sum_{i=1}^n p_i)$. Combining all of the above, we conclude that each assignment of jobs to a machine produces $\bigO\left(n\cdot 2^{2k} \cdot k \cdot \sum_{i=1}^n p_i\right)$ triplets. Therefore, the state space across all $m$ machines has size bounded by $\bigO\left(\left(n\cdot 2^{2k} \cdot k \cdot \sum_{i=1}^n p_i\right)^{m}\right)$. Multiplying this by the cost of computing the value in each cell makes the total complexity of filling the inner DP $\bigO\left(\left(n\cdot 2^{2k}\right)^m \cdot \left(k \cdot \sum_{i=1}^n p_i\right)^{m+1}\right)$, hence concluding the proof.
\end{proof}

\subsection{About the generalization to minimizing the sum of weighted completion times}
\label{app:sumwjcj}

To minimize $\sum w_j C_j$, each DP cell is modified to hold the minimum weighted sum of completion times for its partial schedule. The initial state, where all subsets are empty and all target times are zero, now holds a value of~$0$, while cells corresponding to infeasible states now hold~$\infty$. During the recursive search, instead of looking for at least one \texttt{yes} among the searched cells, the algorithm seeks the minimum value across them. Once the minimum cost among all searched cells has been found, the current cell is assigned that value plus the cost of the newly fixed job, $w_{j_\ell^\star} \cdot t_\ell$ (where $w_{j_\ell^\star}$ is the weight of job $j_\ell^\star$). Finally, when the DP table is filled, the optimal solution is found by searching its last level and returning the minimum weighted sum of completion times.

\section{Missing proof of Section~\ref{sec:limitations_order}}
\label{app:limiations_order}

\thmhardorderbased*

\begin{proof}
        The problem is in \NP, since any proposed schedule can be straightforwardly checked for feasibility. We now reduce the $3$-SAT problem to  $P \mid \mathrm{prec} \mid C_{\max}$ with $2$-perturbed order-based encoding.
\paragraph{Construction.}
        Let $\Phi = (X,C)$ be an instance of $3$-SAT, where $X = \{x_1,...,x_n\}$ is the set of variables and $C = \{c_1,...,c_m\}$ is the set of clauses. We assume, without loss of generality, that each variable occurs at least once as a positive literal and at least once as a negative one.\footnote{If a variable occurs only positively, we can set it to \texttt{True}. If it occurs only negatively, we can set it to \texttt{False}. In both cases, all clauses containing that variable are satisfied and can be removed, and the remaining formula is satisfiable if and only if the original formula is.} We construct an instance of $P \mid \mathrm{prec} \mid C_{\max}$ with $2$-perturbed order-based encoding as follows. For each variable $x_i,\ i\in [n]$, we construct two unit-time jobs, $T_i$ and $F_i$, representing the truth values (\texttt{True} and \texttt{False}). We refer to such jobs as \emph{variable jobs}. Additionally, for each clause $c_j,\ j \in [m]$, containing literals $\ell_j^1,\ \ell_j^2,\ \ell_j^3$, we construct three unit-time jobs $L_j^1,\ L_j^2,\ L_j^3$. We refer to those jobs as \emph{literal jobs}. Furthermore, we introduce two \emph{buffer jobs}, required for the correctness of our reduction, and a set of \emph{gadget jobs} that express the relation between each literal and its variable; both of these will be explained later in the proof.
        
        Each literal of a clause is either a variable or a negated variable. Essentially, our goal is to force the literal job to be preceded by the variable job whose truth value satisfies the literal. This should be done in such a way that a feasible schedule with the target makespan exists if and only if there exists at least one true literal in each clause, i.e. if $\Phi$ is a yes-instance of $3$-SAT. 

         We utilize $3m+1$ machines. We set the target makespan $t=2n+3m+2$ and we will show that~$t$ can only be achieved if there is no idle time on any machine.

        \paragraph{Precedence constraints.}

        We now explicitly state the precedence constraints between jobs. First, the first buffer job, $B_1$, which is a unit-time job, has to precede both $T_1$ and $F_1$. Then, for all $i \in [n-1]$, the variable jobs $T_i,\, F_i$ corresponding to the variable $x_i$ precede both variable jobs $T_{i+1},\, F_{i+1}$ corresponding to $x_{i+1}$. Note that there are no precedence constraints between $T_i$ and $F_i$, so they might be scheduled in any order or in parallel. Similarly, for all $j \in [m-1]$, each triplet of literal jobs $L_j^1,\, L_j^2,\, L_j^3$ associated with a clause $c_j$ precedes the triplet of literal jobs $L_{j+1}^1,\, L_{j+1}^2,\, L_{j+1}^3$ associated with $c_{j+1}$. Again, there are no precedence constraints between $L_j^1,\, L_j^2$ and $L_j^3$, so they may be scheduled in any order or in parallel. All variable jobs precede all literal jobs and all literal jobs precede the second buffer unit-time job, $B_2$.
         
         We express the relation between a literal and the truth value that satisfies it with the use of gadget jobs. Let $\ell_j^l$, with $j\in[m]$ and $l \in \{1,2,3\}$, be a literal that is satisfied when $x_i$ is \texttt{True} (resp. \texttt{False}), for some $i \in [n]$. Their relation is expressed through the following set of gadget jobs:
        \begin{enumerate}
            \item A job $J_{j,1}^l$, with processing time $2i$.
            \item Two jobs that must both succeed $J_{j,1}^l$, with no precedence constraints between them:
            \begin{enumerate}
                \item A job $J_{j,2}^l$ with processing time $2n+1-2i+3(j-1)$. This job must additionally succeed $T_i$ (resp. $F_i$) and precede $L^l_j$.
                \item A job $J_{j,3}^l$ with processing time $1$.
            \end{enumerate} 
            \item A job $J_{j,4}^l$ with processing time $3m-3j+3$. This job must succeed both  $J_{j,2}^l$ and $J_{j,3}^l$. Additionally, it must succeed all jobs $L_{j-1}^l$, for $l \in \{1,2,3\}$, if $j>1$.
        \end{enumerate}
        We often refer to this set of gadget jobs as the gadget jobs that \emph{correspond} to literal job $L_j^l$. Note that the sum of these processing times is exactly $2n+3m+2=t$. Furthermore, notice that the total number of jobs is $2n+2+15m$. Let us make a couple of key observations:
        \begin{enumerate}[label=(\alph*)]
            \item The total processing time of all jobs is $(3m+1)(2n+2+3m)$. Since there are $3m+1$ machines at our disposal, it follows directly that the makespan cannot be less than $t$. Furthermore, a makespan of~$t$ is achievable if and only if there exist no idle times.\label{obs:no_idle}
            \item Since there are exactly $3m+1$ jobs ($B_2$ and jobs $J^l_{j,4}$) that are not required to precede any other jobs, all these jobs must end on the last time step $t$ in order to meet the deadline. Otherwise, there would be fewer than $3m+1$ jobs on the last time step, contradicting observation~\ref{obs:no_idle}.\label{obs:end_no_idle}
        \end{enumerate}

        \paragraph{The predicted order.}
        It remains to construct an order-based encoding $\sigma$ for which there exists a makespan-minimizing schedule whose (order-based) encoding $\sigma^*$ satisfies $|\sigma^{-1}(i) - (\sigma^*)^{-1}(i)|\leq 2$, for all $i \in [2n+2+15m]$. We construct the order-based encoding $\sigma$ as follows:
        \begin{itemize}
            \item The first $2n+1$ positions contain the buffer job $B_1$, followed by the variable jobs in non-decreasing order of index $i \in [n]$, i.e., $B_1,\, T_1,\, F_1, \dots,\, T_n,\, F_n$.

            \item The next $3m+1$ positions contain the literal jobs in non-decreasing order of clause index $j \in [m]$, i.e., $L_1^1,\, L_1^2,\, L_1^3, \dots,\, L_m^1,\, L_m^2,\, L_m^3$, followed by the buffer job $B_2$.

            \item The remaining $12m$ positions are occupied by the $3m$ sets of gadget jobs, appearing consecutively in non-decreasing order of clause index $j \in [m]$ and, within each clause, in non-decreasing order of literal index $l \in \{1,2,3\}$:
            \begin{itemize}
                \item $J^1_{1,1}, J^1_{1,2}, J^1_{1,3}, J^1_{1,4}$ for literal job $L_1^1$, then
                \item $J^2_{1,1}, J^2_{1,2}, J^2_{1,3}, J^2_{1,4}$ for literal job $L_1^2$, then
                \item $J^3_{1,1}, J^3_{1,2}, J^3_{1,3}, J^3_{1,4}$ for literal job $L_1^3$, then
                \item $J^1_{2,1}, J^1_{2,2}, J^1_{2,3}, J^1_{2,4}$ for literal job $L_2^1$, and so on, up to
                \item $J^3_{m,1}, J^3_{m,2}, J^3_{m,3}, J^3_{m,4}$ for literal job $L_m^3$.
            \end{itemize}
        \end{itemize}
        
        It now remains to show that there exists an order-based encoding $\sigma^*$ of a makespan-minimizing schedule such that $|\sigma^{-1}(i)-(\sigma^*)^{-1}(i)| \leq 2$, for all $i \in [2n+2+15m]$. We defer this step to a later point in the proof to first establish how jobs in $\sigma$ are assigned to machines. 
        
        Intuitively, we want the schedule derived from this order-based encoding to assign all buffer, literal and variable jobs on the same machine. Each set of gadget jobs corresponding to a literal job $L_j^l$, for all $j \in [m]$ and $l \in \{1,2,3\}$, is then assigned to its own machine. This yields a structure similar to the one depicted in Figure~\ref{fig:hardness_example}, where a literal job can be scheduled first in its clause if and only if the variable job whose truth value satisfies the literal is scheduled at the earliest possible time. We will prove that a makespan-minimizing schedule can always be obtained by swapping the position between variable jobs of the same variable (displacement of at most $1$ position), literal jobs of the same clause (displacement of at most $2$ positions) or between $J_{j,2}^l$ and $J_{j,3}^l$, for some $j \in [m]$ and $l \in \{1,2,3\}$ (displacement of at most $1$ position). 
    
    \begin{figure}[ht!]
    
    \centering
    \resizebox{\textwidth}{!}{
    \begin{tikzpicture}[
        job/.style={draw, rectangle, minimum height=1cm, align=center, font=\small, fill=white},
        >=Stealth,
        every node/.style={font=\sffamily}
    ]

        \foreach \x in {2, 3, 4, 5, 6, 7, 8, 9, 10, 11, 12, 13, 14, 15, 16} {
            \draw[dotted, gray!70, thick] (\x, 0.5) -- (\x, -10.0);
        }
        \draw[->, thick] (1.5, -10.0) -- (16.5, -10.0) node[right, font=\small\bfseries] {Time};

        \node[font=\large\bfseries, fill=white, inner sep=4pt] at (5.5, 0) {$(x_1 \lor x_2 \lor x_3) \land (\neg x_1 \lor \neg x_2 \lor \neg x_3)$};

        \node[job, minimum width=1cm] (B1) at (2.5, -2.0) {$B_1$};
        \node[job, minimum width=1cm] (T1) at (3.5, -2.0) {$T_1$};
        \node[job, minimum width=1cm] (F1) at (4.5, -2.0) {$F_1$};
        \node[job, minimum width=1cm] (T2) at (5.5, -2.0) {$T_2$};
        \node[job, minimum width=1cm] (F2) at (6.5, -2.0) {$F_2$};
        \node[job, minimum width=1cm] (T3) at (7.5, -2.0) {$T_3$};
        \node[job, minimum width=1cm] (F3) at (8.5, -2.0) {$F_3$};

        \node[job, minimum width=1cm] (L11) at (9.5, -2.0) {$L_1^1$};
        \node[job, minimum width=1cm] (L12) at (10.5, -2.0) {$L_1^2$};
        \node[job, minimum width=1cm] (L13) at (11.5, -2.0) {$L_1^3$};
        \node[job, minimum width=1cm] (L21) at (12.5, -2.0) {$L_2^1$};
        \node[job, minimum width=1cm] (L22) at (13.5, -2.0) {$L_2^2$};
        \node[job, minimum width=1cm] (L23) at (14.5, -2.0) {$L_2^3$};
        \node[job, minimum width=1cm] (B2) at (15.5, -2.0) {$B_2$};

        \node[job, minimum width=2cm] (J11) at (3, -5.0) {$J_{1,1}^1$};
        \node[job, minimum width=5cm] (J12) at (6.5, -5.0) {$J_{1,2}^1$};
        \node[job, minimum width=1cm] (J13) at (9.5, -5.0) {$J_{1,3}^1$};
        \node[job, minimum width=6cm] (J14) at (13, -5.0) {$J_{1,4}^1$};

        \node[job, minimum width=2cm] (J21) at (3, -8.0) {$J_{2,1}^1$};
        \node[job, minimum width=1cm] (J23) at (4.5, -8.0) {$J_{2,3}^1$};
        \node[job, minimum width=8cm] (J22) at (9.0, -8.0) {$J_{2,2}^1$};
        \node[job, minimum width=3cm] (J24) at (14.5, -8.0) {$J_{2,4}^1$};

        \draw[->, rounded corners] (J11.south) -- ++(0, -0.6) -| ([xshift=-0.25cm]J13.south);
        \draw[->, rounded corners] ([xshift=1.5cm]J12.south) -- ++(0, -0.3) -| ([xshift=0.2cm]J14.south);
        \draw[->, rounded corners] ([xshift=+0.25cm]J13.south) -- ++(0, -0.5) -| ([xshift=-0.2cm]J14.south);
        \draw[->, rounded corners] (J11.south) -- ++(0, -0.4) -| (J12.south);

        \draw[->, rounded corners] (J21.south) -- ++(0, -0.4) -| ([xshift=-0.25cm]J23.south);
        \draw[->, rounded corners] (J21.south) -- ++(0, -0.6) -| ([xshift=-3cm]J22.south);
        \draw[->, rounded corners] ([xshift=0.25cm]J23.south) -- ++(0, -0.8) -| ([xshift=0cm]J24.south);
        \draw[->, rounded corners] ([xshift=3cm]J22.south) -- ++(0, -0.4) -| ([xshift=-0.5cm]J24.south);

        \draw[->, green, thick] ([xshift=-0.2cm]T1.south) -- ++(0, -1.2) -| ([xshift=-2.1cm]J12.north);
        \draw[->, green, thick] ([xshift=1.5cm]J12.north) -- ++(0, 1.0) -| (L11.south);
        
        \draw[->, green, thick] ([xshift=0.2cm]F1.south) -- ++(0, -4.0) -| ([xshift=-2cm]J22.north);
        \draw[->, red, thick] ([xshift=2cm]J22.north) -- ++(0, 4.0) -| (L21.south);
        
        \node[align=left, text=red!80!black, font=\scriptsize, fill=white, draw=red, inner sep=3pt, rounded corners] at (12.5, -3.5) {Error:\\$L_2^1$ cannot be the\\first literal job of\\its clause};

    \end{tikzpicture}     
    }    
    \caption{An example of how the literals $x_1$ and $\lnot x_1$ of the $3$-SAT instance $(x_1 \lor x_2 \lor x_3) \land (\lnot x_1 \lor \lnot x_2 \lor \lnot x_3)$ are handled in our construction. For visual clarity, we omit the precedence constraints involving the buffer jobs, as well as those among variable jobs, among literal jobs, and between literal and variable jobs. We also omit the constraints between $J_{2,4}^1$ and literal jobs and the set of gadget jobs that do not express a relation involving the variable $x_1$. Assume that we would like to satisfy the literal $x_1$. The literal job $L_1^1$ can be the first of its clause if $T_1$ is processed before $F_1$. As a result of processing $F_1$ after $T_1$, the gadget job $J_{2,2}^1$ cannot be processed immediately after $J_{2,1}^1$. Hence, we are forced to swap it with $J_{2,3}^1$ to avoid introducing idle times. This prevents the literal job of $\lnot x_1$, namely $L_2^1$, from being the first of its clause.}
    \label{fig:hardness_example}
\end{figure}
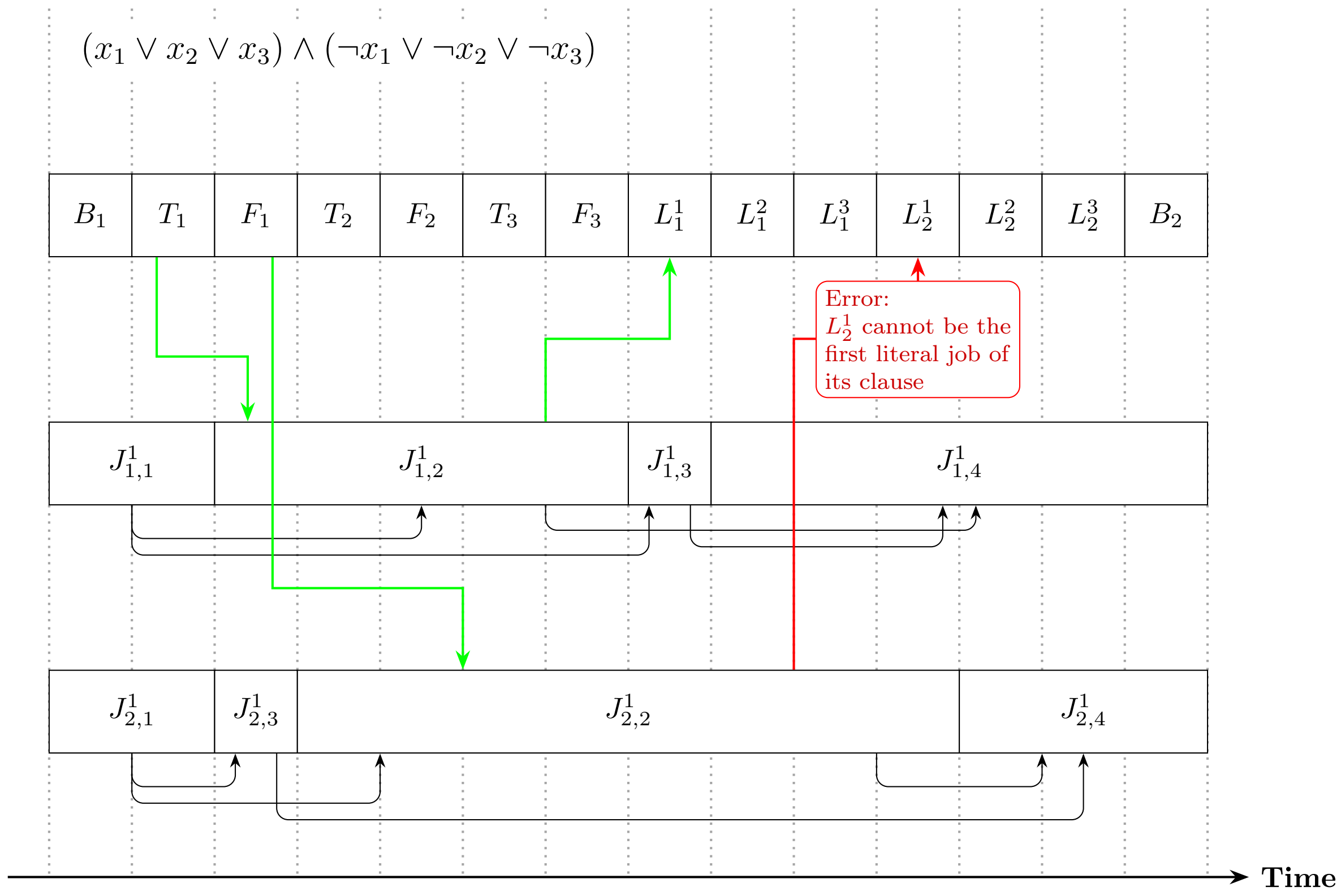

        \paragraph{If $\Phi$ is satisfiable, the target makespan $t$ is achievable.}
        Let $\tau: X \rightarrow \{\texttt{True, False}\}$ be a truth assignment satisfying $\Phi$. We call the first of the $3m+1$ machines the \emph{main machine}. This machine will process the two buffer jobs, as well as all variable and literal jobs. The rest of the machines are called \emph{gadget machines}, since they will only process gadget jobs. We construct a schedule achieving a makespan of $t$ as follows.

        Starting with the main machine, place the buffer job $B_1$ in the first time slot. Then, sequentially schedule all variable jobs (ordered by non-decreasing order of index $i \in [n]$), followed by all literal jobs (ordered by non-decreasing order of clause index $j \in [m]$). For all $i \in [n]$, if $\tau(x_i) = \texttt{True}$, place $T_i$ before $F_i$; otherwise, place $F_i$ before $T_i$. Since $\Phi$ is satisfiable, each clause $c_j$, for $j \in [m]$, contains at least one literal $\ell_j^l$ such that $\tau(\ell_j^l) = \texttt{True}$, for some $l \in \{1,2,3\}$. We place the corresponding $L_j^l$ first among the jobs of its clause (see the following paragraph for why this is feasible), followed immediately by the other two literal jobs in arbitrary order. Finally we put the buffer job $B_2$. This results in the main machine having a load of $2n+2+3m$. Observe that, by the above construction, the variable jobs of $x_i$ will be processed in time slots $2i$ and $2i+1$, while the literal jobs of literals in clause $c_j$ will be processed in time slots $2n+1+3(j-1)+1,\, 2n+1+3(j-1)+2$ and $2n+1+3(j-1)+3$. The latter implies that a literal job is never the cause of a delay for a gadget job $J_{j',4}^l$, with $j'\in [j+1,\,m]$; by the time its predecessor $J_{j',2}^l$ is completed (see the following paragraph), all literal jobs that must precede it have also finished.

        We now move to the gadget machines. Each such machine is assigned a set $J_{j,1}^l,\, J_{j,2}^l,\, J_{j,3}^l,\, J_{j,4}^l$ of gadget jobs, for all $j \in [m]$ and $l \in \{1,2,3\}$. Suppose these gadget jobs express the relation between the literal $\ell_j^l$ and the variable $x_i$, $i \in [n]$, with $\ell_j^l = x_i$. Analogous arguments hold for the case where $\ell_j^l = \lnot x_i$, requiring only to replace each following mention of $T_i$ with $F_i$. Recall that for all $j \in [m]$ and $l \in \{1,2,3\}$, there exists at least one $\ell_j^l$ for which $\tau(\ell_j^l)=\texttt{True}$. In the schedule we construct, the job $J_{j,1}^l$, having a processing time of $2i$, is the first to be assigned to the machine. Then, there are two cases: 
        \begin{itemize}
            \item If $T_i$ is scheduled at time slot $2i$ (meaning $\tau(x_i)=\texttt{True}$), we schedule immediately after $J_{j,1}^l$ the gadget job $J_{j,2}^l$, followed by $J_{j,3}^l$. Observe that $J_{j,2}^l$ will finish at time slot $2n+1+3(j-1)$, allowing $L_j^l$ to be the first job scheduled from its clause. Schedule $J_{j,4}^l$ last on this machine, which results in a load of $2n+2+3m$.

            \item If $T_i$ is scheduled at time slot $2i+1$ (meaning $\tau(x_i)=\texttt{False}$), then $J_{j,2}^l$ cannot be scheduled at the same time slot. Hence, we schedule immediately after $J_{j,1}^l$ the gadget job $J_{j,3}^l$, followed by $J_{j,2}^l$. Notice that $J_{j,2}^l$ will finish at time slot $2n+1+3(j-1)+1$, forcing $L_j^l$ to be either the second or the third job scheduled from its clause. Schedule $J_{j,4}^l$ last on this machine, which results in a load of $2n+2+3m$. 
        \end{itemize} 

        Since every machine has a load of exactly $2n+2+3m$, the makespan is equal to $t=2n+2+3m$ as well. We remark that there exists an order-based encoding $\sigma^*$ of the above constructed schedule for which $|(\sigma^*)^{-1}(i) - \sigma^{-1}(i)| \leq 2$, for all $i \in [2n+2+15m]$. Indeed, variable jobs shift positions by at most $1$ (by swapping $T_i$ and $F_i$), literal jobs by at most $2$ (by swapping $L_j^1$ and $L_j^3$), and gadget jobs by at most 1 (by swapping $J_{j,2}^l$ and $J_{j,3}^l$).

        \paragraph{If the target makespan $t$ is achieved, then $\Phi$ is satisfiable.}
        Assume that a makespan of~$t$ is achievable. We first show that if a schedule admits a makespan $t$, then for all $j\in [m]$ at least one of the literal jobs $L_j^1$, $L_j^2$, $L_j^3$ is scheduled in time slot $2n+1+3(j-1)+1$. 
        
        To this end, note first that by the precedence constraints between $J^l_{j,1}$, $J^l_{j,2}$ and $L^l_j$, the latter cannot begin before time slot $2n+1+3(j-1)+1$.  
        
        Now let us focus on time slots $2n+1+3(j-1)+2$ and $2n+1+3(j-1)+3$. On these time slots: 
        \begin{itemize}
            \item for $j' \in [j]$, the three jobs $J^1_{j',4}$, $J^2_{j',4}$, $J^3_{j',4}$ are being processed. Using observation~\ref{obs:end_no_idle}, they end at time slot $t$, so they occupy all times slots between $t-(3m-3j'+3)+1$ and $t$, with $t-(3m-3j'+3)+1=2n+3j'\leq 2n+3j=2n+1+3(j-1)+2$.
            \item for $j'' \in [j+1,m]$, the three jobs $J^1_{j'',2}$, $J^2_{j'',2}$, $J^3_{j'',2}$ are being processed, since they cannot end earlier than time slot $2n+1+3(j''-1)\geq 2n+1+3(j-1)+3$. 
        \end{itemize}        
        So at least $3m$ gadget jobs are processed on each of these two time slots.

        Now we prove our statement by a simple downward recursion.
        
        \emph{Basis $(j=m)$}: The literal jobs $L_m^1$, $L_m^2$, $L_m^3$ are restricted to the three time slots $2n+1+3(m-1)+1$, $2n+1+3(m-1)+2$ and $2n+1+3(m-1)+3=t-1$ (since $B_2$ succeeds them). There is at most one free spot in each of the latter two time slots, so at least one of the three jobs has to be scheduled on the first of these slots, i.e., $2n+1+3(m-1)+1$.

        \emph{Inductive Hypothesis}: Suppose that at least one of the literal jobs $L_{j''}^1$, $L_{j''}^2$, $L_{j''}^3$ is scheduled in time slot $2n+1+3(j''-1)+1$, for all $j'' \in [j+1,m]$.

        \emph{Inductive Step}: We prove that this also extends to $L_j^1$, $L_j^2$, $L_j^3$.
        Indeed, notice that these three literal jobs must precede the jobs $L_{j+1}^1$, $L_{j+1}^2$, $L_{j+1}^3$, at least one of which is processed at time slot $2n+1+3((j+1)-1)+1=2n+1+3(j-1)+4$, by the inductive hypothesis. Therefore, $L_j^1$, $L_j^2$, $L_j^3$ are restricted to the three time slots $2n+1+3(j-1)+1$, $2n+1+3(j-1)+2$ and $2n+1+3(j-1)+3$. There is at most one free spot in each of the latter two time slots, so at least one of the three jobs has to be scheduled on time slot $2n+1+3(j-1)+1$.
        
        To find a truth assignment $\tau$ that satisfies $\Phi$, we examine the time slots $2n+1+3(j-1)+1$ in our schedule, for all $j \in [m]$. By the above inductive proof, there exists at least one literal job $L_j^l$, $l \in \{1,2,3\}$, scheduled in this exact time slot. This job corresponds to a literal $\ell_j^l$ of clause $c_j$. We construct $\tau$ by setting all such literals to $\texttt{True}$. If a variable is not assigned a truth value by this procedure, we set a truth value to it arbitrarily. Naturally, since at least one literal in each clause is set to \texttt{True}, such an assignment satisfies $\Phi$. It remains only to show that $\tau$ is free of contradictions, by proving that it is impossible for the schedule to force us to assign the same truth value to both a variable $x_i$ and its negation $\lnot x_i$, for any $i \in [n]$. 

        Suppose we have a literal job $L_j^l$, with $j \in [m]$ and $l \in \{1,2,3\}$, corresponding to a literal $\ell_j^l = x_i$ for some variable $x_i$, with $i \in [n]$. For $L_j^l$ to be scheduled at time slot $2n+1+3(j-1)+1$, the gadget job $J_{j,1}^l$ (with processing time $2i$) must start at time slot $1$ and the gadget job $J_{j,2}^l$ (with processing time $2n+1-2i+3(j-1)$) must be scheduled immediately after the former finishes, starting at time slot $2i+1$. For that to happen, the variable job $T_i$ must be scheduled at the latest at time slot $2i$. The same scheduling arguments hold for a literal job $L_{j'}^{l'}$, with $j' \in [m]$ and $l' \in \{1,2,3\}$, that corresponds to a literal $\ell_{j'}^{l'} = \lnot x_i$, meaning that for it to be scheduled at time slot $2n+1+3(j'-1)+1$ the variable job $F_i$ must be scheduled at the latest at time slot $2i$. 
        
        We now show that it is impossible for both literal jobs to be scheduled at their earliest potential time slots, by proving that at most one of $T_i$ or $F_i$ can be scheduled by time slot $2i$. We prove this by induction on the variable index $i \in [n]$, while also proving that none of these variable jobs can be scheduled after time slot $2i+1$. This guarantees that the truth assignment $\tau$ will never assign the same truth value to both a variable and its negation, thus concluding the proof. 

        \emph{Basis ($i=1$)}: Since the unit-time buffer job $B_1$ must precede both $T_1$ and $F_1$, the variable jobs can be scheduled no earlier than time slot $2$. Additionally, if either variable job finishes at time slot $4$, then scheduling a corresponding gadget job $J_{j,2}^l$ immediately after it, followed by $J_{j,4}^l$, for some $j \in [m]$ and $l \in \{1,2,3\}$, admits a completion time of $4+(2n-1+3j-3)+(3m-3j+3)=2n+3m+3>t$. Thus, both $T_1$ and $F_1$ must be scheduled in either time slot $2$ or $3$.

        We now show that it is not possible for both variable jobs to be scheduled at time slot $2$. Recall that each set of gadget jobs has a unit-time slack, hence at time slot $2$ all jobs $J_{j,1}^l$, for $j \in [m]$ and $l \in \{1,2,3\}$, are being processed. Therefore, $3m$ machines are occupied at that time slot, meaning that at most one of $T_1$ and $F_1$ can be scheduled at that time.

        \emph{Inductive Hypothesis}: Suppose that at most one of $T_{i'}$ or $F_{i'}$ can be scheduled by time slot $2i'$ and none of them can be scheduled after time slot $2i'+1$, for all $i' \in [i-1]$.
        
        \emph{Inductive Step}: We prove that this property also extends to $T_i$ and $F_i$. By the inductive hypothesis, it follows directly from the precedence constraints that these jobs can be placed no earlier than time slot $2i$. Additionally, if either variable job finishes at time slot $2i+2$, then scheduling a corresponding gadget job $J_{j,2}^l$ immediately after it, followed by $J_{j,4}^l$, for some $j \in [m]$ and $l \in \{1,2,3\}$, admits a completion time of $(2i+2)+(2n+1-2i+3j-3)+(3m-3j+3)=2n+3m+3>t$. Thus, both $T_i$ and $F_i$ must be scheduled in either time slot $2i$ or $2i+1$.

        We now show that it is not possible for both variable jobs to be scheduled at time slot $2i$. Recall that each set of gadget jobs has a unit-time slack, hence at time slot $2i$ all gadget jobs $J_{j,2}^l$ associated with a variable $x_{i'}$ with $i' \in [i-1]$ are being processed (otherwise, the target makespan is not met). Additionally, all gadget jobs $J_{j,1}^l$ associated with a variable $x_{i''}$ with $i'' \in [i,n]$ are still processing. Therefore, there is only one available machine during time slot $2i$, meaning that at most one of the two variable jobs $T_i$ and $F_i$ can be scheduled then.

        \paragraph{Asserting that $\sigma$ is a $2$-perturbed order-based encoding.} We have proven that if the target makespan $t$ is achievable, then $\Phi$ is satisfiable. For the case where $\Phi$ is satisfiable, we have already demonstrated the existence of an order-based encoding $\sigma^*$ of a schedule with makespan $t$ for which $|(\sigma^*)^{-1}(i) - \sigma^{-1}(i)| \leq 2$, for all $i \in [2n+2+15m]$. It remains to show that if the target makespan $t$ is not achievable, there exists an order-based encoding $\sigma^\star$ of a makespan-minimizing schedule for which $|(\sigma^\star)^{-1}(i) - \sigma^{-1}(i)| \leq 2$, for all $i \in [2n+2+15m]$. To do so, it suffices to prove that there exists such a $\sigma^\star$ that admits a schedule with makespan $t+1$.

        We again utilize the structure of one main machine and $3m$ gadget machines. On the main machine, we schedule the buffer job $B_1$, followed immediately by all variable and literal jobs in the order that they appear in $\sigma$, and finally the buffer job $B_2$. However, this time we leave time slot $2n+2$ of the main machine empty, resulting in a total completion time of $2n+3m+3 = t+1$. Notice that this schedule satisfies the buffer constraints, as well as all precedence constraints between any two variable or literal jobs. It remains to show that no precedence constraints involving gadget jobs are violated either.

        We now move to the gadget machines. Each such machine is assigned the four gadget jobs $J_{j,1}^l$, $J_{j,3}^l$, $J_{j,2}^l$, $J_{j,4}^l$ in this order without idle times, for all $j \in [m]$ and $l \in \{1,2,3\}$. Assume that these gadget jobs are associated with a variable $x_i$, $i \in [n]$, and notice that we always schedule jobs $J_{j,3}^l$ before $J_{j,2}^l$. This schedule, if feasible, admits a total completion time of $t$ for each gadget machine. We prove that this assignment does not violate any of the remaining precedence constraints.

        First, it follows directly that the precedence constraints between gadget jobs corresponding to the same literal are satisfied. Moving to the precedence constraints of $J_{j,2}^l$, notice that on the main machine both $T_i$ and $F_i$ are completed before time slot $2i+2$. Since $J_{j,2}^l$ is scheduled immediately after $J_{j,1}^l$ and $J_{j,3}^l$, it starts at time slot $2i+1+1 = 2i+2$, respecting its precedence constraint. Having a processing time of $2n+1-2i+3(j-1)$, this means that $J_{j,2}^l$ is processed up to and including time slot $2n+3(j-1)+2$. With the added idle time, the literal jobs $L_j^l$ on the main machine start no earlier than $(2n+1)+1+3(j-1)+1 = 2n+3(j-1)+3$. This ensures the constraint of $J_{j,2}^l$ preceding $L_j^l$ is not violated. Finally, the gadget job $J_{j,4}^l$ start at time $2n+3(j-1)+3$, whereas all literal jobs $L_{j-1}^l$ on the main machine are completed no later than time slot $(2n+1)+1+3((j-1)-1)+3 = 2n+3(j-1)+2$. Thus, this precedence constraint is met too.

        Since no precedence constraints are violated, we have proven that there exists a feasible schedule with makespan $t+1$ whose order-based encoding $\sigma^\star$ satisfies $|(\sigma^\star)^{-1}(i) - \sigma^{-1}(i)| \leq 1$, for all $i \in [2n+2+15m]$. Therefore, $\sigma$ is always a $2$-perturbed order-based encoding of a makespan-minimizing schedule.   
\end{proof}

\end{document}